\documentclass[twocolumn,showpacs,floatfix]{revtex4}
\usepackage{graphicx}
\usepackage{dcolumn}
\usepackage{bm}
\begin{document}
\title{ Heuristic model of teaching }
\author{ Stanis{\l}aw D. G{\l}azek } 
\affiliation{ Institute of Theoretical Physics, 
              University of Warsaw, Poland }
\date{4 September 2008}
\begin{abstract}
A simple physical model differentiates 
effective from ineffective teaching and 
identifies elements that make teaching 
productive, with specific implications
concerning training of teachers.
\end{abstract}
\maketitle
\section{ Introduction }
\label{sec:I}

It is reasonable to attempt to describe the
process of teaching in terms of a model because it
is known that models can lead to improvement in
our understanding of natural phenomena and the
understanding may allow us to distinguish useful
from useless action. A physical model proposed in
this article provides a picture in which effective
teaching is clearly separated from ineffective
teaching. The model picture also allows for
identification of elements which lead to teaching
that may be called productive. These insights have
implications for the training of teachers.

Section~\ref{sec:need} discusses the reasons why a
model of teaching is needed and sets the stage for
the next two sections, in which the model proposed
in this article is described.
Section~\ref{sec:kot} describes a kinematical part
of the model that is sufficient to identify the
difference between effective and ineffective
teaching. Section~\ref{sec:dot} describes a
dynamical part of the model that leads to the
concept of productive teaching. Conclusions are in
Section~\ref{sec:c}.

\section{ The need for a model }
\label{sec:need} 

Building models is a time-honored practice to
organize thinking about concepts in physics. For
example, the concept of the atom has been
discussed since ancient times. About a century
ago, Bohr was equipped with enough data on atomic
phenomena to conceive and describe a concrete
image of an atom~\cite{Bohr}. Bohr's model allowed
physicists to focus their attention on the
dynamical issues in the physics of atoms and
gradually replace the model with quantum
mechanics. The latter provided a mathematical
basis for the progress that followed~\cite{Dirac}.

Thus, one can say that physical models may be
useful in two ways. One way is that a model
provides a context in which a concept can be
spoken about in concrete terms. The concreteness
of the context helps eliminate confusion and
allows researchers to focus on the dynamics of the
observed phenomena. Another way is that a model
can be wrong in the sense that it disagrees with
results of experiments or observations. Such
findings provide the basis for seeking a better
next-generation model. The same methodology may be
applied to teaching. There is no reason to limit
this methodology to physics teaching; it applies
also to teaching in other disciplines.

A simple physical model of teaching (MOT, or just
``model'') described in this paper incorporates
the concepts of a teacher and a student through
the concept of observers. Two observers can
communicate with each other about phenomena they
observe and one can help the other in
understanding what happens. In a preliminary way,
the model also allows for incorporation of the
concept of forces that drive this process.
Recognition of the existence of relevant forces,
and how they manifest themselves in a
teacher-student relationship, leads to a
description of the concept of productive teaching.
Since the same model structure appears valid in
all contexts in which the process of teaching may
occur, the scope of the model is not limited to
the case of a teacher teaching a student in a
classroom at a school or university.

The claim that a physical MOT may be formulated in
a simple way requires explanation because it is
known from physics education
research~\cite{Arons,Literature,
McDermottEditorial} that the process of teaching
is not simple. Procedures used by teachers to
specifically check what students actually learn in
a physics course involve complex physical notions
and as such are not simple~\cite{Bao}. One may
expect that a process of teaching is even more
complex than in physics when the subject matter
involves phenomena that cannot be explained simply
in terms of physics. 

From a psychological point of view, the process of
teaching cannot be separated from the process of
learning and the latter is not commonly defined in
simple terms~\cite{Plato,Rousseau,Montessori,
Dewey,Piaget,Sarason,GlazekSarason}. From a
neuroscientific point of view, the process of
learning by the brain is a subject of intense
study, and complex phenomenological and
theoretical models are needed to describe how
individual neurons and networks of neurons
work~\cite{Hebb,Marr,Rakic,NeuralPrinciples,Arbib,
DayanAbbott}. Besides the brain, the process of
learning involves a student's body and the body is
also complex. For example, the involvement of
hands in the process of learning is not
simple~\cite{FRWilson}. The role that hands play
in learning is related to the complex process of
evolution of species~\cite{Darwin}, including the
ability to learn as a means of increasing the
chances of survival and reproduction.
Highly-evolved learning abilities in humans
emerged from processes that correlate the behavior
of individuals with the environment they live in,
and the latter influences individuals in very
complex ways (for a popular explanation of how
advanced human behaviors, such as altruism of
teachers, could emerge in the evolution,
see~\cite{Dawkins}; cf.~\cite{Fehr,Harbaugh}). At
the level of contemporary society, the processes
of teaching and learning in a large system of
education can be seen as exceedingly
complex~\cite{SarasonCulture,Comenius} and may
fascinate physicists~\cite{WilsonDaviss,
Charpak,Wilson2005,Wieman}. The need to corral
the complexity of the process of teaching provides
perhaps the strongest motivation for building a
suitable physical model. Without a MOT, the
purpose of educational reform is ambiguous.

A physical MOT can be simple only in
the sense that the basic elements of the process
of teaching can be identified using physical
concepts such as an observer and a frame of
reference. The complexity of teaching comes from
the complexity of the events that observers
communicate about and from the complexity of the
observers themselves, including the forces that
drive their learning and ultimately determine what
teaching may accomplish. Given this starting
point, a simple model will be obtained in the next
two sections. 

\section{ Kinematical part: effective teaching } 
\label{sec:kot}

In order to kinematically describe what happens in
the process of teaching, one first postulates that
every student has a mental ``dictionary'' that
associates specific word definitions with specific
concepts, such as the sequences of words
associated with ``velocity'' or
``acceleration''~\cite{wordsandconcepts}. A
growing body of research on misconceptions in
physics shows how bizarre the word entries for
these two concepts can be in a student's mental
dictionary~\cite{acceleration1,acceleration2}. The
source of this variety of entries is that every
student's dictionary results from his or her prior
learning history and every student has a different
history.

By definition, in the process of teaching a
teacher communicates with a student and, as a
result, the student's dictionary grows and
improves in accuracy. However, one should remember
that students who take a course have different
mental dictionaries concerning concepts to be
taught, such as in physics, and faculty members
would only deceive themselves if they believed
that every student has the same interpretation for
the language that they use, especially when the
students are only beginning to learn this
language.

It is postulated that the words and concepts
learned by a student are stored in his or her
brain in a coded form that ultimately amounts to
sequences of coordinates (numbers) in a
multidimensional space. For brevity, this space
will be called the student's space of knowledge.
Nothing more needs to be said about this space
except that the model will require an additional
structure to provide a representation of what a
word or a concept means to the student. This
additional structure will be introduced soon. 

In support of the concept of coordinates in the
student's space of knowledge one might recall
mathematical considerations which imply that
statements in any logical system can be
represented in terms of numbers~\cite{Goedel}.
Similarly, neural science suggests that learning
and memory storage processes can be understood in
terms of numbers that represent in specific units
how networks of neurons work, such as voltages,
currents, densities of ions,
etc.~\cite{Arbib,DayanAbbott}. Psychological
studies also have a history of attempts to
describe the meaning of words and concepts to
individuals in terms of numbers~\cite{Osgood}.
Attempts to develop artificial intelligence
presume that the human brain can be understood as
an information-processing machine that ultimately
operates only with numbers~\cite{AI0,AI1,AI2}.

However, the MOT described here postulates the
student's space of knowledge not for the purpose
of explaining, but for the purpose of
circumventing, the issues of complexity and lack
of understanding of the human brain, its
development and function. The goal is to model
concepts of teaching in such basic terms that a
person interested in these concepts can form
suitable entries for them in his or her own
dictionary without the necessity of first engaging
in life-long studies of what is currently known
about the human brain and its psychology. 

Thus, the changes in a student's mental dictionary
are represented in the model by transformations of
the coordinates of words and concepts in the
student's space of knowledge. Teaching a subject
matter to a student is meant to cause the
coordinates of words and concepts pertaining to
the subject matter in the student's space of
knowledge to form the intended relations among
each other, thereby forming an accurate
dictionary. 

It follows that a teacher and a student need to
communicate well with each other in order for the
teaching process to stay on track. The
communication must be good in both directions (not
only from the teacher to the student but also from
the student to the teacher) because the only
person who knows what the student's knowledge
consists of after interpretation using the
student's dictionary, is the student. 

The internal mental processes in the student's
brain enable the student to produce conversations
or questions using his or her own dictionary.
These internal processes involve the coordinates
of the words and concepts in the student's brain
but the student is not conscious of what the
coordinates are and what their numerical values
are. What the student can be conscious of
regarding his or her mental dictionary is, for
example, a sense of confusion about what the
teacher is saying, or a visible disagreement
between a prediction the student makes using his
or her words and concepts about a possible result
of an observation or experiment and a result he or
she actually obtains. 

When the student undertakes a conscious effort to
make a change in his or her space of knowledge and
asks questions, he or she needs to get useful
answers from the teacher, which means answers that
suggest to him or her what needs to be changed and
how. In order to be able to help the student, the
teacher must know what the student thinks. Using
the model, one can say that the task of the
teacher is to establish (as best as one can) the
coordinates of relevant words and concepts in the
student's brain, identify the transformation that
will reduce or remove the confusion that stems
from the improper relations among these
coordinates, and then use the words and concepts
already known to the student to provide him or her
with information about what needs to be
reconsidered in order to deal with the confusion
effectively, the effect being the required change
in the student's dictionary. 

A student may learn without constant help from a
teacher, or people who play this role, directly or
through textbooks and other means. However, by
definition, the model presented here pertains to a
process of teaching that involves a teacher who
helps a student~\cite{externaldictionary}.

Further description of the model concerns a
representation of the meaning of words and
concepts and how the meaning is conveyed from
person to person. Both aspects are incorporated
into the model using the concept of observers and
how they communicate about events.

\subsection{ Students and teachers as observers }
\label{subsec:sto}

Consider a student to be an observer whose brain
registers events and stores information about them
in memory. Every event is registered at some place
and time and has some content. It is reasonable to
postulate that the content of an event registered
in a brain ultimately amounts to a sequence of
parameters. (A concrete way of imagining these
parameters is described in
Appendix~\ref{app:dimensions}.) Essentially, the
parameters describe activity in all parts of the
brain involved in the process of registration. The
parameters are treated in the model as coordinates
in the space of events. If the content of every
event registered in a brain is assumed equivalent
to a sequence of values of $n$ parameters, the
number of dimensions in the space of events is $N
= 4 + n$; 4 for place and time and $n$ for
content. This space of events is used below to
model the process of transfer of information about
the meaning of words and concepts between a
teacher and a student. 

Firstly, it is postulated that the meaning of a
word (or concept) is represented in the student's
brain in the form of memory about a set of events
that are associated in the brain with the word (or
concept). The model is not meant to explain the
neurobiological nature of how the memory functions
so that the association can be formed. The
association is taken for granted in order to
represent the fact that concrete examples allow
people to quickly identify the intended meaning of
words and concepts. 

Secondly, it is postulated that a student creates
and changes his or her own dictionary of words and
concepts on the basis of observation of
correlations between two sets of events: one set
that the student associates with a word definition
and another set that the student associates with a
concept. The identification of these correlations
does not have to be fully conscious. But it is not
necessary to determine the extent to which the
identification of correlations is conscious or not
conscious for the model to provide a picture of
what happens in the process of teaching, including
the transfer of meaning of words and concepts.

In the process of teaching, a teacher introduces a
set of events considered relevant and provides
students with information about them, using words
(defined to a various degree of precision) to
describe the events and related concepts. The
relevant events are, for example, experimental
demonstrations of phenomena. According to the
model, a student's dictionary changes as a result
of the student's attempts to correlate
demonstrated events with their description by the
teacher. The teacher is another observer who helps
the student understand the observed events
using the introduced concepts. The student tries
to improve his or her own dictionary by making it,
from the student's point of view, resemble the one
of the teacher. More generally, teachers and
students become familiar with many events in their
lives and teaching often refers to and certainly
draws on this experience instead of using direct
experimental demonstrations. 

\subsection{ Communication between observers }
\label{subsec:co}

Once the kinematics of teaching is reduced to
communication between observers about events,
further description of the model concerns the
process of communication. 

In physics, effective communication between two
observers (meaningful, precise, and fast) uses
coordinates in their respective frames of
reference. The use of frames of reference is a
hallmark of physics~\cite{Einstein}. Each observer
builds her or his own frame of reference according
to universal rules. Two observers can effectively
exchange information using coordinates of events
as soon as they establish and know (understand and
remember) how their frames are related. The
relationship is found in physics using the
following procedure~\cite{Einstein,classical}.
   
The observers make use of a set of physically
well-defined events that each of them can
unambiguously recognize (in the sense of classical
physics). For example, an event in which two
initially distinct material points meet each other
is a physically well-defined event. The
coordinates of the two points coincide in all
frames of reference when the points coincide.
Having introduced a suitable set of well-defined
events, both observers can find the coordinates of
these events in their respective frames. Knowing
the coordinates in both frames, the observers can
figure out the rules of correspondence between
their coordinates for all events in the set.
Subsequently, they can extrapolate this
correspondence to a general relationship between
their frames of reference that is supposed to
connect their coordinates for all events. They
inspect many cases and, if they gain confidence in
the extrapolated relationship they use, they
consider it valid. 

For example, if the correspondence in the set of
well-defined events is figured out to be a linear
one in an $N$-dimensional vector space, a set of
only $N$ well-defined and linearly independent
events (in terms of their coordinates) is
sufficient to find all parameters of a general
linear relationship between the two frames of
reference. If the required relationship is not
linear, a more complex, case-specific reasoning is
required and a simple, generally valid
relationship is not guaranteed. Note that
Appendix~\ref{app:dimensions} suggests significant
topological complexity of the space of events
registered in a student's brain. The same concerns
the teacher's space of events. A linear
relationship between the frames of reference in
the student's and teacher's spaces of events is
unlikely. 

In the case of teaching introductory physics, the
set of well-defined events may take the form of
tabletop experiments that students carry out with
their own hands and whose results they first
analyze each from her or his own point of view and
in a small group. They then discuss their
observations, confusions, discoveries, and
conclusions with an instructor. Such discussions
create and change the dictionaries students have
for thinking and communicating about physical
phenomena~\cite{McDermottModules,
McDermottTutorials}. 

Since the process of teaching is postulated to
involve the space of knowledge and the space of
events, it follows that the process of
communication between brains requires the
establishment of frames of reference in two spaces
per brain. Indeed, the model associates such two
frames with every brain. Thus every word, concept,
and event becomes equivalent in a brain to some
sequence of co-ordinates, and discussions about
events using all kinds of languages are reduced to
a transfer of sequences of co-ordinates from one
brain to another. Since it is obvious from the
point of view of physics that a meaningful
exchange of sequences of co-ordinates between
observers must be preceded by establishment of a
relationship between their frames of reference, it
is now also clear that the process of establishing
a dictionary must involve the processes of
building frames of reference and finding
relationships between them. 

However, the process of communication between two
brains cannot proceed in as simple and direct a
way as the process of communication between two
observers in physics. The postulated reason is
that we do not have access through thinking to the
values of coordinates that our brains assign to
events they register and to words and concepts we
use in communication about the events. A teacher
may imagine the coordinates and relationships
among them in a student's brain for the purpose of
teaching. But neither the teacher nor the student
knows the coordinates with which the student's
brain registers events or stores words and
concepts. Therefore, one has to accept that the
process of associating words and concepts with
events must exhibit a large degree of ambiguity,
which cannot be eliminated by precise matching and
changing of the numerical values of coordinates of
words and concepts on the basis of directly
measuring and comparing them.

Instead of the direct measurement and comparison
of coordinates, the model postulates that the
creation and development of a dictionary in the
student's space of knowledge is essentially based
on a process of trial and error. A word or concept
is unconsciously assigned its initial coordinates
in the space of knowledge. These coordinates are
associated with the coordinates of concrete events
stored in memory. The initial association can
happen quickly but it is not certain and requires
testing before it is promoted to a more stable
memory. One may postulate that the initial
association depends on the unconscious structural
and functional features of the brain. One is also
free to postulate that the association involves
feedback from complex processes that include both
short- and long-term memory and conscious
thinking. But the nature of the association does
not matter for the model.

Of importance is the postulate that the initial
association between coordinates in both spaces is
modified and stabilized in the brain via the
process of registration of new events and
communication about them. The growth and change in
the brain tissue and its functioning that result
from this process are accepted in the model as an
underlying neurobiological realization of
learning. However, precise knowledge of the
neurobiological realization is not needed to
understand the model. For example, the meaning of
a word is postulated to become eventually
established in the following way.

The coordinates initially assigned to a word in
the space of knowledge are unconsciously
correlated in the brain with coordinates of some
events stored in memory. The initial association
is not entirely random, neither is it precise.
Then the word is used in thinking, absorbing
information, and communication about events. This
process uncovers defects in the initial
association through identification of
misunderstandings~\cite{learning}. The association
is changed to a new one that appears to reduce the
confusion and is used in further thinking,
experimenting, observation, and communication.
Eventually, subsequent uses of the word cease to
lead to confusions that require modification of
the association and the meaning of the word is
obtained in the form of a stable association
between the coordinates of the word in the space
of knowledge and the coordinates of corresponding
events in memory. A similar, multistage process is
postulated for concepts. Identification of the
meaning of concepts in terms of events precedes
the introduction of their names in the dictionary.

Ultimately, the rank and order in the space of
knowledge is postulated in the model to be brought
about by observing events, and communicating about
them with other observers. Thus, a language is
built and learned, and the words in it and
concepts required to understand and predict events
are tuned with increasing accuracy. 

Whether one communicates about current events or
events that are only remembered, or merely
imagined on the basis of memory of the actual
events (such as the motion of a point along a
straight line, Einstein's gedanken experiment
\cite{Wertheimer,EinsteinHadamard,Hadamard}, or an
event described in a book or shown in a film), is
of secondary importance. The same scheme is
expected to apply to teaching in all disciplines,
not just physics. 

\subsection{ Effective teaching }
\label{et}

What follows is the model description of the
process referred to as effective teaching. For the
purpose of differentiating between effective
teaching and another process that may be referred
to as ineffective teaching, the latter will be
presented first. This order of presentation brings
out the conceptual contrast between the two
processes, providing an example of the utility of
the model.

Consider a teacher, for brevity called $T$, and 
a student, for brevity called $S$. $T$ is supposed 
to teach $S$.

In the case of ineffective teaching, $T$ tries to
elicit from $S$ a reproduction of a sequence $t$
of words (or other symbols) that $T$ presents to
$S$. Eventually, $S$ reproduces $t$, from memory,
but without having much of an idea why and what
$T$ wanted to convey by $t$. The sequence $t$ may
be similar to some sequences that $S$ already
knows, but whether these sequences correspond to
what $T$ meant by $t$ and what was the goal of
talking about $t$ remains unclear to $S$. $S$ will
soon forget the sequence $t$ because there is no
reason to remember it.

Such ineffective teaching does not include the
process of establishing the meanings of words and
concepts that $T$ uses and $S$ is supposed to
learn about. In the model picture of ineffective
teaching, what $T$ does is, in essence, supplying
$S$ with a sequence of co-ordinates valid in the
frame of reference of $T$ without trying to find
out how the frame of reference of $S$ is related
to the frame of reference of $T$. Then $T$ expects
to obtain the same sequence back, but has no way
to establish what this sequence means in the $S$'s
frame of reference. Thus, the model makes it
obvious that the process of ineffective teaching,
based on transferring information only one way, is
fundamentally inadequate. At the same time it
becomes clear that ineffective teaching is
characterized by arbitrary assumptions about how
$S$'s brain associates meaning with words and
concepts in terms of events. It is assumed that
$S$ does it in the same way as $T$ so that the
dictionaries of $T$ and $S$ are the same. $T$'s
dictionary is assumed to be well-calibrated with
respect to the world, including $S$ as a part of
the world. It is also assumed that $S$ can be
considered well-taught if he or she repeats the
statements that $T$ wants him or her to repeat.
This is what is ordinarily described as rote
memorization.

The model thus predicts that ineffective teaching
leads to a lack of communication between $T$ and
$S$. It generates confusion and correlates word
definitions with names of concepts in the mind of
$S$ in a disordered fashion, devoid of meaning. On
the surface, things appear not that bad but only
because the ambiguities of thinking and language
hide the imperfections of the communication. 

Further, the model implies that when $T$ interacts
ineffectively with a whole group of students
instead of just one, most of them are confused and
each of them in a different way. Having no chance
to understand what is going on, the students lose
interest. $T$ is not able to regain the students'
attention and disciplinary measures are likely to
dominate other aspects of teaching. An educational
system may be unable to identify dysfunctional
situations of ineffective teaching and students
may be required to pass tests. They are then
judged on the basis of the numerical values of
their scores. The problem is that these values
reflect not whether the students learn but how the
system functions. 

In the case of effective teaching the situation is
remarkably different. According to the model, the
initial task of $T$ is to establish how the frame
of reference of $S$ is oriented with respect to
the frame of reference of $T$. This means that $T$
first tries to find out what statements $S$
already knows regarding the subject of study and
what $S$ means by them, i.e., what events
correspond to these statements according to $S$.
$T$ uses these events to estimate how the
dictionaries of $T$ and $S$ are related to each
other. Further tuning of communication between $T$
and $S$ involves hands-on activity during which
concrete events are associated with concrete words
and concepts. 

When key parts of the dictionaries of $T$ and $S$
are already tuned, which is described in the model
by saying that the relation between $T$'s and
$S$'s frames of reference is approximately known
to $T$ and $T$ knows how to use the dictionaries,
$T$ becomes able to try to understand what to do
in order to convey information contained in the
sequence $t$ to $S$ in terms of sequences that $S$
already understands and can think and talk about
with confidence. Eventually, $T$ finds the
sequence $s$ that conveys to $S$ the same
information about the world that the sequence $t$
corresponds to in the dictionary of $T$. If $s$
represents some important insight, $S$ will use it
and not forget. This experience paves the way for
further steps in building $S$'s dictionary with
$T$'s help. $T$ can always arrange a dialogue
around issues that are comprehensible to $S$, and
the dialog helps $T$ identify issues that still
confuse $S$. 

In the case of a group of students, a discussion
is led by $T$ in order to identify a set of issues
that all members of the group find unclear. In
terms of the model, this discussion allows $T$ to
approximately understand orientation of the
students' frames of reference and co-ordinates of
entries in their dictionaries. This understanding
allows $T$ to help the students in beginning a
coherent study of the identified issues. Thanks to
the meaningful communication, $T$ is welcomed by
students to play the role of an advisor and as
such can transparently influence the course of
study so that it may reach its stated goals.
Teaching of this kind usually leads students to
discoveries of new aspects of the subject matter.
Students may subsequently consider the new aspects
important and learn more about them. Students may
also realize the value of the process of clear
communication about the subject matter, which $T$
enables them to practice. 

\subsection{ Training in effective teaching }
\label{subsec:tet}

Teachers can incorporate effective teaching into
their work with students if they are trained to do
so. Both pre-service and in-service teachers need
training. The training must itself be an example
of effective teaching so that the trainees can
experience the process first from the standpoint
of students. The resulting appreciation of the
great value of effective teaching in comparison
with ineffective teaching, the value explained
here in physical terms using the MOT, motivates
teachers to incorporate effective teaching into
their work with students. 

McDermott and the Physics Education Group at the
University of Washington have developed a program
for training science
teachers~\cite{McDermottEditorial,
McDermottModules,McDermottTutorials}. The model
concept of effective teaching suggests the
possibility of similar elements also being useful
in training in other disciplines. Indeed, one can
compare effective teaching as defined by the MOT
with the teaching of actors that was described by
Stanislavski in 1936~\cite{Stanislavski} (his
three-volume textbook has had more than 40
editions since then). Such a comparison may
initially appear pointless to a reader focused on
physics education. But, when equipped with the
model, the reader will find that the same elements
are present in both cases, even though they
concern different subject matters and different
dictionaries of words and concepts. Another
example, in which the same elements of teaching
can be identified by a trainee equipped with the
model, is provided by the program of
Clay~\cite{MMClay0,MMClay1,MMClay2}. Clay's
program concerns teaching young children who have
extreme difficulty learning to read and write (it
is now in use in thousands of schools).

The point of these comparisons is that the MOT
allows trainees to notice common aspects of
teaching in all disciplines. This in turn allows
them to properly identify the elements that they
could otherwise misinterpret (associate
exclusively with their specialty) if they thought
that such elements occur only in their discipline.
MOT implies that teaching in all disciplines
requires the establishment of relationship between
the frames of reference of $T$ and $S$ before a
meaningful communication between them can begin.
In all disciplines, $T$ must understand $S$'s
initial vocabulary before $T$ becomes able to help
$S$ in building a proper dictionary. Thus, MOT
predicts that the same basic elements must be
present in training of teachers in all
disciplines.

A chief example of a universal feature is that
training requires time. It must extend over the
period that the trainees need to work out hands-on
examples which enable them to discover the
meanings of relevant words and concepts. The
required amount of time can only be decided by
research. Training that does not provide the
required time cannot accomplish the
mission~\cite{DeweyMontessori}. MOT makes these
predictions physically obvious.

For illustration, consider the case of in-service
science teachers. The minimal period of training
required for acquisition of physical concepts of
mass, force, energy, electric current, and
magnetic field, can be estimated as not shorter
than about two months of study every day of the
week~\cite{Discovery,Hebb1}. At least a year of
further study should follow in which trainees
attempt to implement effective teaching in their
work at school. 

The implementation will encounter difficulties.
For example, one can now predict that it will be
impossible for a science trainee to achieve
success with students if the time allocated in a
school program for teaching the words and concepts
of physics is much shorter than the time required
by a typical human brain for learning these words
and concepts in an effective way. Note, however,
that this and other~\cite{Singham} inconsistencies
between training in effective teaching and known
practices of schooling are not specific to the
subject matter and stem instead from the way the
educational system is organized. This brings the 
development of MOT offered here to the key point 
where it becomes clear that the process of training 
of teachers cannot be considered complete unless it 
addresses the issue of success with students and 
what is meant by the success.

The issue of a trainee's success or failure in
implementing what he or she learns in the process
of training requires an extended discussion. The
discussion begins here and extends into
Section~\ref{sec:dot}, where it proceeds to the
issues of dynamics of teaching and leads to the
concept of productive teaching.

When a teacher steps in front of a class of
students and begins to teach, the concept of
effective teaching in the teacher's brain is soon
confronted with challenging, unavoidable
questions: Do the students want to learn what they
are taught? What if they don't? Should they be
forced? What will the students actually learn by
going through an obligatory course if they are not
interested in the subject and all they want is a
passing grade required for graduation? Will they
ever appreciate the subject if they are not
interested in it and do not want to learn? How to
find out what they really think? What is the
teacher to do? The training of teachers is
incomplete if it does not address these questions.
These questions pertain to teaching in all
disciplines and are not specific to any one
subject matter. 

A trainer of teachers is confronted with similar
questions. This is exemplified by replacing the
words: teacher, student, and passing grade
required for graduation, with the words: trainer,
trainee, and career incentive, respectively. The
resulting questions lead to a new one: How to
design a training program, including selection of
candidates to be trained and trainers to train
them, so that a large percentage of trainees will
afterwards achieve success in working with their
students? The question is important because
teachers cannot consciously engage in a long-term
process of improvement of teaching in schools
unless they are successful with students. Such
success confirms for them that they know what to
do and how to do it, and the clarity of purpose
enables them to continue their work every day.
Without success with students, they burn out.

That a MOT should address the questions stated
above is also suggested by results of a
longitudinal study of the same program that was
used earlier to estimate the minimal period of
training for science teachers~\cite{Discovery}. In
addition to other findings of interest to
educators, the study found that a year after
completion of the training in basic physics almost
three quarters of the program participants eagerly
responded to mail surveys concerning their
subsequent experiences in teaching. However, only
about one third responded after four years. It was
not found why the number dropped so much and the
drop could have occurred for multiple reasons. On
the other hand, teachers' commitment to enhance
the quality of their work can be sustained and
reinforced for many years if they achieve success
with students. One may not exclude that the drop
in the number of responses occurred because the
majority of trainees were not sufficiently
motivated by the results of their work with
students to stay in touch with and provide data to
the program that trained them.

Since the concept of effective teaching does not
answer the questions of practice with students, it
is postulated that the process of training
teachers of all kinds and levels misses something
essential if it is limited to the concept of
effective teaching. A MOT would be most useful if
it could help in the identification and inclusion
of the missing elements. A heuristic point of view
toward this goal is discussed in the next section.

\section{ Dynamical part: productive teaching }
\label{sec:dot} 

Section~\ref{sec:kot} describes the process of
teaching as the sequence of changes in a student's
dictionary that a teacher helps the student make
in order to enable the student to interpret events
that belong in the subject area, and communicate
about them in terms of a language suitable for the
subject. Effective teaching corresponds to the
well-defined sequence that is described
kinematically in Section~\ref{sec:kot}. It
includes establishment of how the frames of
reference of $T$ and $S$ in their spaces of
knowledge and events are related to each other,
establishing what are the co-ordinates of relevant
words, concepts, and events, which are used in
communication, and establishing new entries in the
relevant dictionaries on the basis of observation,
experiment, and exchange of information.

In the context of training of teachers,
Section~\ref{subsec:tet} shows that the
kinematical concept of effective teaching is not
sufficient to address the issue of teachers'
success in working with students. Students respond
to teaching in various ways and a model of
teaching must be able to incorporate the dynamics
of the teacher-student relationship. This section
discusses the dynamics. The discussion addresses
the issue of success with students and leads to
the concept of productive teaching. 

\subsection{ Extension of the dictionary }
\label{subsec:ed}

The dictionary of words and concepts in the space
of knowledge (of a student or a teacher)
introduced in Section \ref{sec:kot} is limited to
the subject matter. For example, in the case of
physics, the dictionary includes words such as
``velocity'' and ``acceleration'' or concepts such
as ``mass'' and ``force.'' But
Section~\ref{subsec:tet} shows that the concept of
teaching involves words and concepts such as
``success with students'' or ``wanting to learn,''
and these suggest a host of other related words
and concepts that are all absent in the
dictionaries for specific subjects such as
physics. Therefore, a complete MOT must include an
extended dictionary that by definition includes
specific dictionaries for all disciplines and the
additional words and concepts that pertain to the
process of teaching irrespective of the
discipline. This extended dictionary will be
called the meta-dictionary of teaching. The
meta-dictionary is required to discuss the
dynamics of teaching. The following examples
illustrate what kind of words and concepts must be
included in the meta-dictionary (again, $S$ means
a student and $T$ a teacher).

When $T$ is to teach a subject matter that is
important for $S$ to know but $S$ does not
understand why he or she needs to know the
subject, is not interested in learning, and does
not want to do what $T$ tells $S$ to do, it is
clear that there exists a barrier in communication
between $T$ and $S$ through which $T$ must break.
The key step is to explain to $S$ how the subject
to be taught is related to $S$'s future (this will
be clarified within the model picture in
Sections~\ref{subsec:hvtof}
and~\ref{subsec:tipt}). The dictionary of the
subject itself is of marginal importance in this
step because $S$ does not know it yet. $T$ must
instead refer to the concepts and words that $S$
already knows, understands, and appreciates
regarding $S$'s own future. This is why the
meta-dictionary needs to contain the concept of
{\it starting from where $S$ is} in a broader
sense than in effective teaching, where the
corresponding concept is limited to starting where
$S$ is only with respect to the dictionary of a
particular subject matter. The broader concept
means that $T$ must assess and utilize a variety
of dictionaries that already exist in $S$'s space
of knowledge. Unless $T$ starts from where $S$ is
in this broader sense and explains why the subject
to be taught is important for $S$, it is likely
that $S$ will not be interested in learning the
subject. 

A primary example of obligatory courses that
students may be not interested in is mathematics.
The lack of appreciation is reflected in their
performance. The percentage of students performing
at required levels drops down from grade to grade
(from above 60\% in 4th grade to below 50\% in
12th grade), being lowest at the time of
graduation from high school~\cite{testreport}. The
drop occurs despite the fact that results of math
tests count toward admission to college. Recent
publicly-discussed data on students' performance
in mathematics in grades 3 to 8~\cite{nyt1,nyt2}
reflect the same trend. The percentage of students
who score at or above proficiency levels drops
from about 80\% in third grade to below 60\% in
eighth grade. If the drop is interpreted as a
consequence of students losing interest in what
math is about and what it is needed for, one has
to conclude that teaching of math is
inadequate~\cite{inadequate}. The inadequacy is
not describable in terms of the language of
mathematics alone. One needs an extended
dictionary that can be used in discussing the
relationship between mathematics and students,
students and their future in society, and thus
also mathematics and society. 

When $S$ thinks that teaching is not adequate, he
or she should be able to say so. For example, it
may be unclear to $S$ why knowledge of the
mathematical concept of integration is required.
$T$ must respond to $S$'s concern using the
dictionary that already exists in $S$'s space of
knowledge. Note, however, that an entirely new
concept emerges. $S$ may be afraid to ask about
the purpose of lessons on integration due to a
fear of consequences if the question is
interpreted as unintelligent or violating the
school order. Such fear prevents $S$ from learning
what it means to speak up on matters that concern
$S$ while the way the teaching process is
conducted influences the development of $S$'s
brain and determines what $S$ knows and
understands about the meaning and significance of
the taught subject. Thus, the way $S$ is treated
in the process of teaching determines $S$'s
competence and self-image as a member of society.
The apparently narrow problem with teaching math
is ultimately related to the concept of {\it
educating a citizen}, which includes experience of
social control, democracy, freedom, and critical
thinking~\cite{DeweyExperience,PiagetFuture,
Whitehead,Korczak,Korczak2,AronsCitizenry}.

These aspects are not specific to any area of
study and play an important role in the dynamics
of teaching (see next sections). They require a
dictionary of words and concepts that are
fundamental to the well-being of society. The
corresponding meta-dictionary of teaching is much
richer than the dictionary-to-be-taught concerning
just mathematics, or any other discipline taken
out of its relevant context. In the case of
mathematics, $T$ may help the class understand not
only why integration is taught but also why $S$'s
question about the purpose of lessons and the way
$T$ answers this question are all important.
Subsequently, the obligatory lesson of mathematics
and the role of $T$ will appear justified instead
of arbitrary from the point of view of $S$ and the
class.

Of course, students will not voice doubts about
the adequacy or purpose of the process of teaching
unless they have a feeling of {\it safety} with
$T$. Otherwise, the fear of negative judgment by
$T$ and hence uncertainty of the future will
silence students. But to gain the {\it trust} of
students and achieve openness in communication
with them so that real issues on their minds have
a chance to get resolved, $T$ needs to think in
terms far broader than the subject matter alone.
The corresponding meta-dictionary of teaching must
contain concepts of safety and trust in
communication between people, in addition to the
words and concepts limited to the dictionaries of
a specific subject and judgment of students'
progress in learning the specific subject.

Safety and trust are important for asking
questions about the subject matter. Asking a
question that discloses confusion may imply that
$S$ does not know or does not understand something
in a situation where others appear to already know
and understand. Such exposure could put $S$ at a
disadvantage in an already stressful situation.
Without feelings of safety and trust, $S$ may
refrain from asking questions.

Teaching does not have to become inadequate and
misguided in purpose if $T$ starts from where the
students are and takes advantage of their {\it
curiosity}. Curiosity is a central concept in the
meta-dictionary of teaching. 

A curious $S$ eagerly confronts a problem to be
solved. Finding it difficult, $S$ is glad to
receive help from $T$, and appreciates $T$'s
contribution. In order to contribute to the
self-motivated study, $T$ must be interested in
and capable of helping $S$ in the learning
process. The self-motivated learning is hard for
$S$ because it involves unlearning, a change in
brain structure that was in place, and learning
anew, creating a new structure. This requires that
$S$ overcomes the stress that is associated with
the changes and performs the work that is needed
to make the changes happen. Curiosity can take $S$
through the difficulties. So, $T$ should sustain
$S$'s curiosity, a condition valid irrespective of
the subject of study.

\subsection{ Inclusion of internal events }
\label{subsec:ie}

Effective teaching concerns a dictionary of a
concrete subject matter and uses events observed
by $S$ and $T$ as the vehicle that conveys between
them the meaning of words and concepts in the
dictionary. For example, $S$ and $T$ discuss
well-defined events on a laboratory table.

A complete MOT contains the meta-dictionary
described in Section~\ref{subsec:ed}. This
dictionary extends beyond the subject matter
itself and includes ``new'' words and concepts
such as success with students, wanting to learn,
starting from where the student is, educating a
citizen, safety, trust, and curiosity. One can
also talk about attitude, attention, engagement,
etc. According to Section~\ref{sec:kot}, in order
to identify the meanings of these words and
concepts, one should associate them with concrete
events. However, the required events are
definitely not of the kind that happen on a 
laboratory table.

In the case of new words and concepts, the events
that matter happen inside $S$ or inside $T$ (see
below), instead of outside of them, as is the case
with the events on a laboratory table. Therefore,
the space of events that is needed in a complete
MOT must include events inside an observer. Such
events will be called internal. Their description
in the model is provided in a few detailed steps
below. The same details will be useful later in
the discussion of the concept of productive
teaching. More precisely, they will be used in the
model to show that the concept of productive
teaching is not merely an addition to the concept
of effective teaching but underlies the latter as
a foundation of the whole process of teaching.

Firstly, consider $T$ teaching $S$ some motor
skill, such as riding a bike, catching or throwing
a ball, or pressing strings on the fingerboard of
a violin; or a mental skill, such as the ability
to focus, be patient, contain emotion, or perform
a gedanken experiment. The process actively
involves $S$ and $T$ as performers of the skill
and thus involves phenomena that happen inside $S$
or $T$ in ways not reducible to learning a
dictionary for description of events outside
observers. Namely, $S$ must focus on what happens
within the body and mind of $S$ when performing a
skill. $T$ judges $S$'s performance from outside
but focuses on what happens within the body and
mind of $S$. Moreover, the analysis of $S$'s
performance that $T$ makes is based on what $T$
knows is happening in the body and mind of $T$
when performing the same skill. It should be
clarified that learning a skill includes a buildup
of a dictionary concerning the skill. One cannot
fully comprehend concepts and words used in
communicating about the skill unless one is able
to perform the skill in a way corresponding to the
meanings that count.

Secondly, consider the model's definition of the
space of events registered in a brain that is
provided in Appendix~\ref{app:dimensions}. The
definition makes it clear that the events
registered in a brain are determined not only by
physical events outside an observer but also by a
multitude of processes that happen concurrently in
the observer's brain. These processes depend on
the neurophysiological history and state of the
brain as a central organ of a living person, which
includes how the brain functions using the senses,
memory, and thinking, both subconsciously and
consciously. It is clear that many physical events
outside an observer are not registered in the
observer's brain because the input they provide is
below the threshold for altering the concurrent
activity in the brain.

Thirdly, while registration of events is
associated with changes in an observer's brain, a
change in the functioning of the observer's brain
is not necessarily expressed in the observer's
overt behavior. If it is not overtly and clearly
expressed, the change cannot be registered and
unambiguously interpreted by another
observer~\cite{testing}. 

In essence, using the model of
Appendix~\ref{app:dimensions}, one can say that
events registered in a brain can be approximately
divided into three classes according to their
dominant coordinates. Some events have dominant
coordinates associated with what is happening
outside of an observer. These will be called
external events. Other events have dominant
coordinates associated with what is happening
inside of the observer. These will be called
internal events. The third class contains events
that have their dominant coordinates associated
both with what happens inside and outside of the
observer. For brevity, these will be called
engaging events. 

In terms of the model nomenclature, one can say
that the concept of effective teaching involves
external events, such as an event on a laboratory
table; engaging events, such as those that cause
changes in $S$'s dictionary concerning what
happens on a laboratory table; and internal
events, such as thinking about an event on a
laboratory table, experiencing confusion regarding
the subject, or formulating questions about it.
The concept of effective teaching ignores and
obscures all internal events that are not directly
related to the improvement of $S$'s dictionary for
the taught subject. 

The ignored internal events include, for example,
wandering of thoughts away from the taught subject
toward issues of main interest to $S$, or $S$'s
plain feeling of aversion toward learning the
subject. The possibility that $S$ sees the subject
as having no relevance to $S$'s way of life and
future is also overlooked. Phenomena such as
boredom, a sense of violation by enforcement of
classroom lessons or homework, feeling insecure in
the presence of $T$, or not trusting $T$ enough to
share important thoughts or information with $T$,
are not included in the model picture of effective
teaching. All internal events that contribute to
(positive) attitudes and manifest themselves in
$S$'s attention or engagement, are considered
given, or stimulated and reinforced by effective
teaching, although no reason is identified in the
kinematical part of the model for the necessity of
their occurrence. Internal events that contribute
to the phenomenon of curiosity are not explicitly
considered. Instead, curiosity for the subject
matter is assumed to be always in place, as it by
definition is assumed to be in place in the case
of $T$, who is also assumed to be interested in
teaching the subject to the students. In summary,
the kinematical concept of effective teaching in
the model does not explicitly include a host of
feelings, thoughts, attitudes, wants, and
expectations that are important for the course and
outcome of the process of teaching.

The above examples make it clear that the concept
of effective teaching described in
Section~\ref{sec:kot} omits the internal events
that provide meaning to the meta-dictionary
described in Section~\ref{sec:kot}. Of course,
ignoring the internal events does not change the
fact that the behavior of $S$ hinges on them and
the process of teaching cannot be successful if it
is incompatible with the internal events. A
realistic MOT must account for their significance.

\subsection{ Communication about internal events }
\label{subsec:cie}

The issues that require the meta-dictionary of
Section~\ref{subsec:ed}, cannot be discussed
between $T$ and $S$ without communication about
internal events. But the communication cannot
proceed according to the scheme of effective
teaching described in Section~\ref{subsec:co}.

In effective teaching, the dictionary of words and
concepts concerning a subject of study is built,
used, and changed with the help of a process of
communication based on a set of well-defined
events; both $T$ and $S$ observe and discuss these
events as two different observers who use two
different frames of reference to describe the same
phenomena. Such a set of well-defined events does
not exist for communication between observers
about events that happen inside one of them. The
difficulty is that, by definition, only one
observer has access to what happens in an internal
event. 

Since the standard methodology adopted in
effective teaching does not apply, the question
arises how $T$ and $S$ may proceed. The model
asserts that the situation is not hopeless because
$T$ and $S$ may rely on their own internal events
in concrete situations and each can try to find
similarities in what the other says regarding
these examples.

The model's description of what may happen in a
dialog between $T$ and $S$ when they attempt to
establish orientation of their frames of reference
with respect to each other, including internal
events~\cite{internal}, is provided below in two
parts. The first part describes only what may
happen and does not address the issue of why the
dialog may proceed in the described way. The
second part, which contains a heuristic point of
view toward the origin of forces that can keep the
dialog on track, is provided in the next
subsection.

Suppose the observer of an internal event is $S$
(an event happens inside $S$). $T$ can infer what
happens but only from the overt, nonverbal
behavior of $S$, and from the verbal information
provided by $S$ in terms of $S$'s dictionary. In
order to decode such information, $T$ must
hypothesize about what happens inside $S$ and
about the meaning of the words $S$ uses. $T$ must
conduct the dialog using and expressing the
hypotheses so that the dialog can lead to
improvement of the hypotheses, or to making new
ones. However, {\it it is up to $S$ whether the
required dialog with $S$ will proceed.} 

In particular, the dialog's prospects depend on
how $S$ perceives $T$. $S$ will judge how $T$
comes across according to the rules chosen by $S$,
not by $T$. The judgment will include the utility
of the relationship between $S$ and $T$ for the
purpose of dealing with the internal events in $S$
by $S$ (see next subsection). No quick fix or
superficial verbal assurance can change the
judgment that $S$ builds over time. Instead, overt
behavior of $T$ in response to overt behavior of
$S$ is registered in the brain of $S$ and this
extended process contributes to $S$'s judgment of
$T$.

If $S$ judges $T$ as helpful, the dialog will
proceed because $S$ expects a benefit. If $T$ is
judged as not helpful, $S$ will not be motivated
to cooperate. These regularities in behavior are
observed in countless
examples~\cite{GlazekSarason}. 

The key to $S$'s willingness to cooperate with $T$
is that $T$ comes across as somebody who wants and
is able to help $S$ unfold the potential that $S$
believes himself or herself to have; potential
that is not reducible to any subject matter or
dictionary. Inclusion of this regularity in
communication between $T$ and $S$ about internal
events is a challenge for every MOT. It will be
addressed in the present MOT in the next sections
(including relevant references). The goal is to
capture in the MOT that $T$'s authentic respect
for the potential of $S$ to develop as a learning
human being and the interest of $T$ in $S$ having
a chance to realize his or her true potential are
both seen and appreciated by $S$ in the process of
building up their interpersonal relationship. Eo
ipso the cooperation happens voluntarily on both
their parts. Only a conscious process supported
from both sides allows $T$ and $S$ to build the
required meta-dictionary. This mutually supported
process is identified in the model as the way
around the basic difficulty of an absence of
well-defined events: wanting allows the
participants to continue despite
misunderstandings. This is also why the
significance of safety and trust was stressed in
Section~\ref{subsec:ed}, even though every normal
human being is able to identify the significance
of safety and trust in their experience.

Since the task of establishing meaningful
communication between two brains is at the
discretion of the brains that are unique as
individuals, and since this task depends on the
interpersonal relationship between them and the
context of their communication that are not fully
predictable, it cannot be carried out in steps
enforced by a {\it predetermined curriculum}
according to a {\it rigid schedule}. The magnitude
of complexity involved in communication between
two brains about internal events boggles the mind
and requires study. In fact, science is yet to
discover a precise methodology~\cite{EBWilson} to
deal with this complexity because the standard
procedure based on well-defined events is not
available. The fact that the present MOT
eventually produces such clear conclusion in
disagreement with most common educational
practices, is an example of the power of models
that is needed in theory of teaching according to
Section~\ref{sec:need}.

This means that teaching involves an art of
communicating despite ambiguities that cannot be
systematically eliminated by any simple procedure.
Performance of this art requires extensive
preparation and continuous training such as
performance in other
disciplines~\cite{SarasonArtofTeaching}. The issue
of required teacher selection and training is
taken up in Section~\ref{subsec:tipt}.

Even if the buildup of the meta-dictionary depends
on the internal events, which depend in turn on
the interpersonal relationship between $T$ and $S$
and complex contexts of their interactions, all of
these elements being unpredictable to such an
extent that teaching cannot be reduced to simple
procedures, the model must now provide a dynamical
idea for what drives the process of teaching in
the right direction. The next subsection
incorporates the required concepts of ``dynamics''
and ``right direction'' into the model.

\subsection{ A heuristic view toward the origin 
of forces }
\label{subsec:hvtof}

Challenges for a MOT stated in the previous
sections must be approached with the realization
that the advanced processes that go on in the
brain cannot be ultimately explained today in
terms of their physics, strictly speaking, because
they are too complex and too little is known and
understood about them. Approaches of a far
less-precise nature than physics, such as Freud's
attempts~\cite{Freud} to understand psychology and
Hebb's idea~\cite{Hebb2} of explaining the same in
terms of neurophysiology, are still far away from
reaching conclusion~\cite{Kandel}. Moreover, all
these approaches concern just one brain. The
process of a teacher teaching students is of a
much finer kind because it involves at least two
brains in interaction in a complex environment and
it causes complex changes in them in complex ways.

But the current lack of fundamental understanding
of the forces relevant to the MOT does not imply
that such forces do not exist, that one may
succeed in teaching by acting against them, or
that one has a chance to seriously tackle problems
of education being ignorant of their role and
strength. The situation is like one with the force
of gravity that causes stones to roll only down a
slope irrespective of whether one understands the
law of gravity or not. 

This section employs the model picture of teaching
developed in the previous sections to describe a
heuristic point of view toward the origin of the
relevant forces. The model view suggests that the
same basic forces always determine the course and
outcome of the process of teaching and that
attempts to teach in ways that act against these
forces cannot lead to success. As a result of
recognizing the depth of the relevant forces'
origin, and thus also their strength and
significance, the next subsection will introduce
the concept of productive teaching.

To begin with, the concept of ``force'' as a
dynamical agent in the process of teaching should
be distinguished from other known concepts that
are associated with the word ``force'' in
teaching. For example, there exist forces that act
in the daily practice of teaching and do not
directly refer to the buildup of dictionaries
through interactions between brains. This concept
of forces includes systemic rules mandated by law
or economic and social pressures that act on
teachers, students, and other stake-holders in the
system. Such forces qualify as constraints or
influences on what happens in schools (and other
places where teaching occurs) rather than causes
of the process of teaching that the model is
about. 

Another important concept is that the force of
teaching is the teacher. In a trivial way, this
may be interpreted to mean that the teacher is the
person who forces students to learn, no matter how
it is done. An opposite meaning associated with
the concept of teacher as a force of teaching is
that the teacher is the identifier, facilitator,
creator, and guardian of contexts in which
students learn~\cite{SarasonCPL1,SarasonCPL2}. A
vast difference between these two interpretations
demonstrates the magnitude of ambiguity that needs
to be eliminated. A more subtle ambiguity is
contained in popular statements that students must
realize that ``only they can learn'' and that ``it
depends on them how much they learn,'' which seem
to imply that the teacher plays only a secondary
role. On the other hand, it is also said that
teachers need to ``create a desire to learn'' in
students. The latter statement can be confronted
with the fact that most children are eager to
learn. Evidently, there is a need for
clarification of the concept of forces in the
process of teaching. 

Consider first a few examples which show that the
action of forces relevant in teaching manifests
itself in recognizable psychological phenomena.
These phenomena may lead to unproductive outcomes
of teaching in major ways even though the subject
has been learned by $S$ sufficiently well to pass
some test. Consider that $S$ has questions
concerning his or her self-image in the context of
learning a dictionary. For example, if $S$ feels
judged as ``not gifted enough to learn physics''
and never talks about it with $T$, $S$'s entire
life may later be tainted by efforts to overcome
the feeling of inferiority. The other highly
unproductive outcome would be that $S$ resigns to
the feeling of inferiority and never becomes a
fully developed person because of the internal
feeling of failure. An opposite case of concern is
an overconfident $S$ who will have to deal with
consequences of serious errors resulting from the
illusion of competence. A common psychological
effect of school or college is that alumni avoid
contact with subjects that were forced upon them
during their academic training. This effect
negatively influences the perception of arts and
sciences in society.

To be more specific, it is known that test anxiety
in students has major implications for their
learning and for measurements of what they
learn~\cite{anxiety1,anxiety2,anxiety3}. In an
environment that stimulates test anxiety, many
individuals cannot learn effectively and their
actual knowledge cannot be reliably assessed. Of
course, the anxiety is not invented by testers:
some natural inner forces lead to anxiety in some
conditions. Although the forces that cause anxiety
are not precisely known, it is known that anxiety
has devastating consequences for teaching. If,
instead of being ignored and blindly caused to
backfire, the same underlying dynamics were better
understood and taken advantage of in the process
of teaching~\cite{Tsystem,Cutts}, it would be
possible to create and maintain conditions that do
not stimulate the destructive anxiety in students
who are prone to it, and do not induce a whole
spectrum of unproductive consequences of such
anxiety in their future behavior. One could employ
the dynamics to instead stimulate the same
students' curiosity about the world and useful
dictionaries, with an entirely different spectrum
of consequences in their future behavior. 

In order to locate forces in a potentially
complete physical MOT (even if the forces are not
fully understood), one needs to delineate a
plausible way in which known physical processes
might in principle be responsible for the
existence of these forces. Once room is made for
these processes in the model structure, and thus
the basic (i.e., following from physical laws)
origin of the dynamics of teaching is incorporated
(even if not directly because of the absence of
required details), a model definition of the
concept of productive teaching will be offered. 

In the model, $T$ is considered an observer who
communicates with $S$ as another observer of the
same world. There is no postulate in the model of
any fundamental difference between observers.
However, $T$ is a more experienced observer than
$S$ and has two special features. One feature is
that $T$ wants to possess expertise about the
world's mechanisms and share it with $S$. Another
feature is that $T$ knows what to do in order to
share this expertise with $S$. Among the
mechanisms of the world, $T$ understands and
appreciates the role of personal growth in human
life. On the other hand, $S$ needs and wants $T$'s
help in learning about the world because $S$ is
convinced that $T$ possesses the expertise $S$
wants to gain and $S$ sees that $T$ wants and
knows how to share it with $S$.

A heuristic point of view toward the physical
origin of these features of $T$ and $S$ involves
several observations put together. These
observations concern an individual and a large
number of generations of many individuals. There
is nothing new about these observations except for
their inclusion into the description of a physical
MOT. The purpose is to make the
heuristic point of view as concrete as possible
within as small an amount of text as possible.
From a vast relevant literature, only a few
references are selected that seem to most
accurately put the observations in the model's
context.

The first observation is that humans learn eagerly
from the moment of birth. Children are curious,
explore, and ask questions of adults. That adults
are motivated to explore is seen in their tendency
to travel, especially after they retire and are
free to make choices about how they spend time.
Apparently, humans need to explore to feel well
because this is the only way available to them for
discovering regularities of the environment in
which they live. They need to know these
regularities to successfully operate in the
environment. Similarly, all anomalies (news) catch
their attention because humans need to figure out
the new elements in order to know how to deal with
them. A brain must feel pleasure when exploring,
uncovering regularities, and explaining surprises,
in order to develop strategies for survival and
apply them in an ever-changing and unpredictable
environment. As suggested already by
Hebb~\cite{Hebb3}, the feeling of pleasure coming
from exploration, observing regularities, and
solving puzzles, can be associated with ``directed
growth or development in cerebral organization.'' 
 
The second observation is that a human being is
the product of evolution~\cite{Darwin}. This means
that the ``growth or development in cerebral
organization'' in a human brain and the associated
feeling of pleasure are a result of a chain of
physical events that took many millions of years.
A huge amount of information accumulated in this
chain of events is encoded in the capabilities of
the genetic material of
individuals~\cite{Dawkins2}. In the course of the
life of a single individual, the individual's
genetic code is expressed in varying conditions in
different ways to various
degrees~\cite{KentrosAttention,Udo}. This process
results in the learning brain. Interactions
between different brains are critical to the
continuation of the chain because they are
involved in the transfer and survival of the
genetic material. Ultimately, the formation of
genetic codes and their expression in processes
that produce functioning human brains happen
according to the dynamical laws of
physics~\cite{quantum}. But there is no simple
path that connects the basic laws of physics to a
single living human brain. Instead, a human brain
is built and functions in the ways that result
from and are informed by an unimaginably large
number of intermediate structural and functional
steps that involve simpler structures involved in
simpler functions. As a result, many layers of
complexity are superimposed in the brain
dynamics~\cite{RGanalogy,RG}. An ultimate
understanding of the brain is a fascinating
prospect and one of, if not the greatest,
challenges to science.

The second observation means that the forces which
determine how brains function and interact result
from and are tested by such a long ``battle of
life''~\cite{Darwin} that they will not yield to
any artificial concept of teaching. Instead,
teaching must be adjusted to them. For example,
there is no point in artificially suppressing the
role of feelings because of the assumption that
they only ``jumble'' rigorous thinking, or in
artificially suppressing rigorous thinking as
``secondary'' to the essence of
humanity~\cite{Snow}. In fact, feelings and
thinking are coupled into an inseparable whole.
One cannot argue that logic is more important than
emotion or that emotion is more important than
logic. The only form of the human brain that
passes the test of evolution involves both.
Attempts at artificial separation will encounter
resistance since only a whole brain can function
as a learner and a teacher. 

The third observation is that teaching (such as
parents' teaching of children, or teaching of less
experienced group members by more experienced
members) helps individuals develop values in
behavior and knowledge, including the development
of language~\cite{Darwin}. This is why individuals
can grow far beyond what any one of them could
ever accomplish from scratch or separately. The
ultimate recursion is that the increasing
capability for teaching using increasingly
advanced language allows us to describe the world
we live in to new generations with increasing
precision. The new generations are thus equipped
to achieve greater scope, accuracy of knowledge
and understanding than previous ones, achieving
greater capability to take action.

In the context of dynamics of the model of
teaching developed here, these three observations
are interpreted as inescapably suggesting that the
origin of internal events which manifest
themselves as wanting to learn, wanting to teach,
and wanting to take action, is of the same depth
as the origin of man. Therefore, it is postulated
that the dominant internal events in the process
of teaching are related to who $S$ and $T$ are and
may become as human beings. Each has a strategy
for this purpose. The evolutionary perspective
makes it obvious that disclosure of such a
strategy to a stranger is associated in a brain
with danger. On the other hand, the disclosure may
lead to success thanks to cooperation. Both $S$
and $T$ judge and choose the depth of
communication that suits them. 

In the process of teaching, when $S$ and $T$
encounter difficulties in communication due to
ambiguities in their respective dictionaries and
when they need to share information about internal
events in order to explain what happens, they
operate near the limits of communication that they
set previously as useful and safe. Pushing these
limits, with some necessary elements of curiosity,
pleasure, and risk~\cite{Hebb4}, is what breaks
new ground in the process of teaching according to
the present model. The concepts of a ``right
direction'' and ``dynamics'' of this process can
now be included in the model.

The process of teaching is essentially
``blind''~\cite{blind} in the same sense, in which the
underlying laws of physics and the resulting
process of evolution are blind, and it is driven
by the same forces that drive evolution. But since
it is the highest existing form of transfer of
information from generation to generation
(actually, from a brain to a brain) and as such
includes the transfer of values that allow humans
to operate at the level we have currently reached,
many characteristics of teaching appear to have a
purpose. This is viewed as a misconception
analogous to suggesting that cows have milk so
that people may milk them.

Instead, the model postulates that the concept of
``right direction'' can be associated with the
increase of options that individuals in principle
have for choosing their own steps into the future
as their understanding of the world increases.
(Examples from modern history that illustrate the
complexity of the phenomena that contribute to
this process can be found in
Refs.~\cite{RosenbergBirdzell,Hughes,Bennett}.)
But ``in principle'' does not imply that every
individual is equally aware of the possibilities
and limitations, has equal access to the resources
and equally contributes to their replenishment,
can benefit from the development in the same way
as others and provides others with equal services,
etc. The concept of ``right direction'' that $T$
may adopt in teaching $S$ is defined in the model
by saying that $T$ chooses steps that lead $S$ to
greater awareness of possibilities and
limitations, more efficient access to resources
and participation in their replenishment, wiser
consumption of benefits and more competent service
to others, etc. Most succinctly, $T$ and $S$ share
an understanding of their world and values. 

The concept of ``dynamics'' which causes $T$ and
$S$ to make concrete choices in interacting with
each other is reduced in the model to the
statement that the phenomenon of teaching is
shaped through evolution. It takes forms
corresponding to the qualities and knowledge of
its participants and how they communicate. Most
succinctly, $T$ and $S$ cooperate with each other
in agreement with their needs and understanding of
the world and values. 

As a result of gathering enough experience and
insight, $T$ becomes consciously aware of the
process of creation and transfer of knowledge and
values, and realizes that this process includes
$T$'s own communication with students. From that
moment on, $T$ can consciously work on better
understanding how to communicate with students and
transfer to them foundations of the meaning of
what $T$ has comprehended, so that the values and
knowledge the students learn in the process help
them move into their futures. 

Self-consistency of the model dynamics requires
that the process of buildup and transfer of the
meta-dictionary of teaching is perpetual. This
means that the transfer happens in agreement with
the existing values and knowledge and is carried
on and evolves as these imply, from generation to
generation. So, through participation in the
process, $T$'s students become familiar with the
mission $T$ serves. Appreciation of $T$'s work by
$S$ signifies that $T$'s mission is accomplished
in the case of $S$. 

That the model dynamics may resemble what
currently happens only in a minority of schools
(some examples are described in
Refs.~\cite{Bensman,Levine,Mortenson}) is a
separate issue of great practical significance. An
in-depth discussion of this issue is beyond the
scope of this article. But if the process of
teaching in every school and the model picture of
teaching were similar today, it would mean that we
have already entered the era of {\it consciously}
taking advantage of the mechanisms of our origin
in {\it designing} conditions of our further
development. In fact, as of yet we have not.

\subsection{ Productive teaching }
\label{subsec:pt}

Teaching may be called productive when it includes
all the elements of effective teaching found in
Section~\ref{et}, uses the meta-dictionary of
teaching to solve the problems presented in
Section~\ref{subsec:tet} (as described in
Sections~\ref{subsec:ed}-C), and proceeds as a
result of action of the forces discussed in
Section~\ref{subsec:hvtof}. 

In other words, the process of productive teaching
is not blindly enforced by $T$ but stems and
branches from the will of $S$ with the help of
$T$. $T$ helps $S$ understand what is relevant and
motivates $S$ to participate in making the
teaching process happen in terms of engaging
events. In particular, $S$ feels free to exchange
information about internal events with $T$ (and
other members of the group when more people are
involved). This freedom results from and
contributes to the buildup of an interpersonal
relationship between $S$ and $T$ and becomes
palpable along the establishment of the
meta-dictionary that they use to solve problems
they encounter.

While effective teaching discussed in
Section~\ref{sec:kot} is a concept that may be
tied to some subject of study (and measured solely
in terms of knowledge and understanding of the
specific subject), productive teaching produces a
person who finds pleasure in learning and wants
and knows how to learn more, where the word
``more'' includes future subjects that are not
known and not predictable by $T$ during the
process of teaching. No matter what subjects are
covered in the process of productive teaching, $S$
learns to use the inner forces of learning that
manifest themselves in $S$'s curiosity, wanting to
learn, and joy, even if it gets hard to make
progress. The open state of mind of $S$ that
results from productive teaching is not measurable
in terms of knowledge or understanding of the
subject matter that $S$ is taught. One may say
that productive teaching concerns teaching a
person, not a subject matter. A person thus
educated knows the forces that drive learning and
how to use them.

In particular, the new insights, and observation
of many instances of $T$'s helpful behavior in the
interpersonal relationship with $S$ (and with
others as witnessed by $S$), allow $S$ to realize
that the approach of productive teaching, adopted
by $T$, is driven by the respect that $T$ has for
$S$ as another human being in need of learning.
This instructive interpersonal relationship,
created by $T$ and perceived by $S$ as a fruitful
one, shows $S$ the necessity and value of $T$'s
work~\cite{withoutT}.

Appreciation of $T$'s work by $S$ is what
completes the definition of productive teaching
according to the model. The outcome of productive
teaching is not only the mastery of subjects and
skills but also appreciation by $S$ of the value
of understanding among people, how it happens and
bears fruits. Through this appreciation and
understanding, productive teaching prepares $S$
for further learning and making choices for
action.

The concept of productive teaching is closely
related to the concept of context of productive
learning that was introduced by
Sarason~\cite{SarasonCPL1,SarasonCPL2} but so far
has not been discussed in terms of a model. Using
the model, one can now say that productive
teaching proceeds through employment of contexts
of productive learning. The insight provided by
the physical MOT is threefold: 1) effective
communication between $T$ and $S$ requires
comprehension of the relationship between the
frames of reference in their spaces of knowledge
and events, which implies that $T$ and $S$ have to
start from finding out what this relationship is
and learning how to use it, 2) the internal events
determine the course and outcome of the teaching
process and $T$ and $S$ need to communicate about
them in order to keep the teaching process on
track, and 3) teaching becomes productive when it
is driven by the forces that characterize $T$ and
$S$ as human beings in pursuit of their goals. One
can now also say that the context of productive
learning is a context in which issues of
importance to $S$ are the center stage and subject
of effective communication with $T$ so that the
three elements identified in the model are present
and $S$ can learn with the help of $T$. This means
that teaching begins with and proceeds in
concrete, real-life contexts that $S$ and $T$
identify together and in which $T$ facilitates
$S$'s learning, creating opportunities for $S$ to
discover new concepts and guarding $S$'s process
of learning from derailment. In the process of
teaching, $T$ also learns: constantly studying how
students learn, building a model of how to teach,
and seeking validation for the model, aiming at
improvement.

Documented examples that illustrate the concept of
a context of productive learning, including its
long-term
consequences~\cite{Bensman,Levine,Mortenson}, also
help illustrate in practical terms what kind of
teaching is described by the physical MOT offered
here. In particular, they indicate what may happen
in the dialog between $S$ and $T$ (see
Refs.~\cite{Levine1,Mortenson1}). Examples that
illustrate the same concepts and can be seen in
popular cinema, are described in
Refs.~\cite{GlazekSarason, GlazekSarason1}. An
example from a program of teaching
science~\cite{Discovery} based on
Ref.~\cite{McDermottModules}, can be found in
Ref.~\cite{GlazekSarason2}.

A few examples given below further illustrate what
typically happens in the context of productive
learning and should occur in practice of productive
teaching according to the model. $S$ thinks it is
best to say, ``I do not know'' when he or she does
not know how to find an answer to a question or
solve a problem. $S$ is confident that this is the
best way to respond when missing a point because
$S$ knows that the shortest path to learning is to
take advantage of communication with $T$ and other
people in the process of filling gaps and seeking
connections. Learning is impeded by hiding gaps in
knowledge and pretending to understand. Similarly,
if a personal image and opinion are more important
than actual learning, disclosure of shortcomings
in knowledge or understanding is out of the
picture. 

In the process of productive teaching, $S$ has no
reason to be afraid to say, ``I do not know,''
because $S$ knows that $T$ understands what such a
statement means, listens to $S$ carefully, and
helps. So, $S$ does not hesitate to say, ``Please
explain this and that because I do not
understand'' when he or she is confused, or, ``I
do not remember'' when memory fails, or, ``Why do
you say such and such?'' if what $T$ says does not
sound clear to $S$, etc. 

In order to appreciate how difficult it is to
create and sustain contexts of productive learning
(especially by the unprepared for the unprepared),
these apparently simple examples can be compared
not only with what happens today in many
classrooms, but also with how people communicate
in other contexts. Consider the contexts of
discussion between a teacher and a principal, a
principal and a superintendent, an employee and a
boss, or a citizen and a government official. In
these and other contexts, the subject matter can
suddenly become secondary in importance to the
issue of dominance of one person's position in the
system. At this point, communication about the
subject matter breaks down and there is no room
for communication about internal events. Instead,
the only course of action for the person in a
weaker position in the system is to adjust to the
decision of the person in a stronger position in
the system. In different circumstances, the
hierarchy of their positions may change and a
reaction based on memory takes place.

In the process of productive teaching, there is no
need for blind measures of discipline and
judgment. Instead, the values transferred in the
process bond $S$ to the idea of participation.
Testing of $S$ by $T$ is essentially replaced by
gathering of information by $T$ about what $S$
does while learning. This information is the basis
for advice that $T$ gives to $S$ so that $S$ can
improve the process and reach the intended
results~\cite{ClayCredo} (see
Section~\ref{subsec:tipt}). 

In summary, one may say that productive teaching
differs from effective teaching by the dynamical
condition that the relevant sequence of changes in
$S$'s dictionary occurs as a result of the action
of the natural forces of learning in $S$.
Therefore, $S$ learns what the word ``learn''
means and becomes a learner for the rest of his or
her life. 

Learning in agreement with the natural forces that
$S$ is equipped with as a human being, allows $S$
to discover the meaning of words and concepts such
as human rights, law, democracy, achievement, and
other entries of fundamental value in $S$'s
dictionary. For example, since the rules of
communication between $T$ and $S$ are not decided
solely by $T$, the issues of motivation and power
in the process of teaching get discussed between
$S$ and $T$. Decisions are made in ways that do
not subject $S$ to the will of $T$. Instead, $S$
is engaged in making decisions in the contexts
familiar to $S$. If $T$ must decide because $S$
does not have the required knowledge, $T$ must
also explain to $S$ why $T$ is responsible for
making the decision, and why $T$ considers a
particular decision to be the right one.

The model described here helps in identifying
several meanings of the word ``productive'' in
connection with teaching. For example, {\it many
correct entries} are produced in the multiple
dictionaries concerning subject matters that
belong to $S$'s space of knowledge. The large
number may be achieved because the process of
teaching takes advantage of the forces that drive
learning and they accelerate it to large speeds.
Since the entries are correct and understood, they
are {\it useful} in thinking about new problems
and become {\it stable} (are not forgotten, as
useless information typically is). When teaching
takes advantage of self-motivated learning, it
produces entries in $S$'s dictionary that are {\it
basic} to the well-being of $S$ as an individual,
such as {\it inquiry, discovery,} and {\it study}.
By learning how to use inner forces, a person may
become {\it creative}. A {\it rich}
meta-dictionary is produced in $S$ for
communication about internal events. This
dictionary contains entries for the {\it values}
that are fundamental to $S$'s future as a member
of a society, indispensable for $S$ becoming a
conscious {\it citizen}. The teaching produces a
person who carries these concepts and values on
{\it in relation to other people}. The overarching
meaning is that the teaching is productive when it
contributes to the student's future as a fully
developing person. This short statement is itself
a sequence of words whose meaning cannot be
grasped without a proper dictionary.

\subsection{ Training in productive teaching }
\label{subsec:tipt}

Training in productive teaching is more difficult
and time consuming than training in effective
teaching (see Section~\ref{subsec:tet}) because
productive teaching requires the buildup of many
words and concepts in the meta-dictionary of
teaching on top of a dictionary of specific
discipline(s). However, in a system in which basic
competence of trainees in a subject matter is
itself obtained as a result of productive
teaching, the difficulty and time consumption can
be predicted considerably smaller than in a system
in which a basic familiarity with the same subject
matter is obtained by the trainees without the
first-hand experience of productive teaching
(first as students, and only then as teachers).
Let the productive system be called $P$, and the
questionable one $Q$.

In the case of $P$, teachers are trained in the
subject matter in ways that create entries in the
meta-dictionary in parallel to the creation of
entries in the dictionary for a discipline. In the
case of $Q$, the dictionary for a discipline
remains uncorrelated with the meta-dictionary.
Moreover, the dictionary for a discipline is
plagued with wrong entries, such as the illusion
that science results from pure logic, the false
impression that art results only from emotion, the
mistaken belief that interpersonal relationships
are irrelevant to the development of science, and
the invalid assumption that the context of
discoveries and creations is not important for how
they occur. $Q$ compounds the difficulties of
training in productive teaching because the
trainees first have to unlearn what was inculcated
in them during their training in the subject
matter. Then they can learn anew. But it is much
harder to change an existing structure in a brain
than to build a new one from scratch.
Unfortunately, the $Q$-like systems greatly
outnumber $P$-like systems (see below). 

In a system of the dominant type ($Q$), almost all
practical suggestions for training in productive
teaching originate one way or another in a small
number of examples that have already been
mentioned (some only through references) during
construction of the model in previous sections. A
few new sources are included in this section as
particularly relevant to the training of teachers.
One can say that this section illustrates that a
model, which organizes thinking about teaching,
leads also to a selection of suggestions for
training. 

The model principle of beginning from
establishment of relationship between frames of
reference and starting where the trainees are
implies that the events suitable for beginning a
buildup of the meta-dictionary of teaching,
irrespective of the discipline, should involve a
subject matter that is familiar to the trainees.
The more familiar the subject matter the easier it
is for a trainee to focus on the issues of
teaching rather than on the issues of the subject
matter~\cite{Wieman1}. For example, consider a
physicist to be trained in productive teaching (a
professional in another discipline could just as
easily be considered instead of a physicist). Let
the subject matter be reading. Instead of reading
Shakespeare or Goethe, however, consider the case
of teaching young children who have difficulty
learning to read. Imagine a visit by the physicist
(as a trainee in productive teaching) to a
training session for teachers of reading in a
program called {\it Reading
Recovery}~\cite{MMClay2,RR}. The physicist
witnesses how the trainees observe what happens
between a child learning to read and a veteran
{\it Reading Recovery} teacher who helps the child
overcome the difficulties the child encounters.
The required setup includes a sound-proof room
with a microphone and a one-way mirror so that the
trainees (and the physicist) can hear from a
loudspeaker and observe through the mirror what
happens in the room between the child and teacher
without being seen or heard by the child. 

Since the physicist and all trainees already know
and understand what it means to read, they are
able to focus on what the teacher does in response
to the difficulties that the child encounters. The
whole group observes how the teacher works with
the child and they discuss what they see, among
themselves and with an instructor who observes the
reading room with them and explains what happens
when the trainees have questions or confusions
arise in their discussion. Using the model, one
can say that the observed events allow the
trainees to build entries in their
meta-dictionaries of teaching without
unnecessarily focusing their attention on the
subject matter, which would impede the process of
building the right entries.

The physicist sees that the teacher behind the
mirror is focused on finding out what blocks the
child in reading a story composed of a few simple
words. The teacher discusses with the child the
meaning of the individual words and the whole
story, illustrated with a picture, and helps the
child understand how to self-correct mistakes the
child makes~\cite{selfcorrectioninmath}. It
becomes clear to the physicist that what the
teacher does is not focused on some abstract
concepts of reading but on discovering how to help
a unique child solve a concrete problem. Once the
child finds a solution, the teacher uses it as an
example and helps the child understand how to seek
solutions to similar problems with concrete
letters, words, or groups of words, explaining
through practice the role of comprehension of the
text and the methods of self-correction. This is a
complex interpersonal process, immersed in a
context in which the child wants to learn to read.

The physicist also observes the behavior of the
teacher trainees and their instructor as they
follow the lesson through the mirror and
loudspeaker. It becomes clear to the physicist
that the instructor and trainees do not talk about
meanings of the read words or sentences. These are
understood well and do not require special
attention. They talk about concepts that matter in
helping the child succeed in becoming a learner.
This is a very complex issue, incomparably more
complex than the simple sentence the child tries
to read. The other clear lesson is that in order
to grasp the meaning of words and concepts in the
meta-dictionary of teaching one must observe
events such as those behind and in front of the
one-way mirror.

The next suggestion about training in productive
teaching concerns reading documents that
illustrate what productive teaching can
accomplish. Consider the experiment carried out by
Schaefer-Simmern in the nineteen
forties~\cite{Henry}. He demonstrated in a number
of cases ranging from mental defectives to people
in business and the professions that one can teach
a person to see art as experience~\cite{DeweyArt}.
His students unfolded their creativity in visual
art. Schaefer-Simmern concluded from his
experiment with mental defectives
that~\cite{Selma}: the experiment ``seems to
confirm the fact that creative activity in the
visual arts can be unfolded and developed in
mentally defective persons to a degree analogous
to that of their mental potentialities. Real
difficulties appeared usually only with
individuals with higher IQ's who had previously
received art instruction based on copying. With
such a background, feeble-minded persons cling to
a technique and slavish imitation which are in no
way related to their stage of visual
comprehension. Nevertheless, the imitative,
memorized picture seems to give them a certain
security. Any attempt to lead such persons back to
their own stage of visual conceiving is usually
resented vigorously because of their anxiety over
losing that sterile mental possession and being
thrown back into a state of uncertainty.'' The
above is only one of many lessons from
Schaefer-Simmern's experiments. Dewey stressed in
his foreword to Ref.~\cite{Henry} that
Schaefer-Simmern's experiments provided an
``effective demonstration of what is sound and
alive in theoretical philosophies of art and of
education.'' Trainees in productive teaching may
be asked if they see parallels between
Schaefer-Simmern's experiments and what they think
is possible to achieve in their disciplines.

Other examples of literature that reports on the
practice of productive teaching in schools are
Refs.~\cite{Bensman,Levine} (see
Section~\ref{subsec:pt}). Trainees in productive
teaching can observe exemplary ongoing programs in
action and learn from individuals working in these
programs. Experienced individuals can play the
role of mentors to the trainees. The trainees may
learn from their mentors how the latter achieved
understanding of productive teaching with the help
of their own mentors of a previous generation.
What were the breakthrough events that helped the
mentors get on their paths to productive teaching?
How long did it take and how did it happen that as
a result of these events they learned how to
create and sustain contexts of productive learning
for their students? Such contexts are certainly
not limited to the classroom in the cases
described in Refs.~\cite{Bensman,Levine}. A
challenge to training programs is identifying
people who can be mentors, and the schools that
are suitable for study and accessible to the
trainees~\cite{Shady}. Currently, the concept of
apprenticeship in productive teaching as a common
path to professionalism in education can only be
considered a concept of the future~\cite{Lortie}.
What the physical MOT provides here is a logical
structure which explains why the above examples of
recommendations for training are not merely items 
in another trade book in education but a necessity 
for building a culture of a meaningful communication 
between teachers and students as learning observers 
of the world. 

In the current circumstances, one of the primary
objectives of training is that the trainees learn
the difference between the classroom regularities
in systems of type $Q$ and $P$. In $P$, students
may study the neighborhood of the school and spend
a lot of their time outside the school building,
and in $Q$ they may almost always stay in the
building. But the trainees need to know and
understand the concept of regularities in the
process of teaching. Some regularities will agree
with the MOT, and some will not.

An example of a study of classroom regularities
that qualifies for incorporation in a training
program in productive teaching is Susskind's study
of question asking~\cite{Susskind1,Susskind2}.
Susskind found that teachers often think that they
ask approximately the same number of questions as
students do and that they expect that only a
slightly larger number of students' questions
would be better. In fact, teachers ask about 25
times more questions per period than all students
in a classroom. They receive less than one fifth
the rate of students' questions that they estimate
as occurring and as desirable. The bottom line is
that on average a student asks one question per
month in all his or her social studies classes
combined (assuming four 45-minute social studies
lessons per week and realistic estimates of time
available for asking questions), whereas teachers
ask about one question per minute. The main
reasons for students to not ask questions are
identified by Susskind to be yelling by teachers
and laughing by peers. Students appear to assume
and adjust to a rule that only teachers are to ask
questions, while the duty of students is to
provide answers.

Susskind reported on rates of asking low-level
questions (related to memorization, i.e., of the
type: Who, what, where?) and high-level questions
(such as concerning causes of wars) by teachers.
He observed that these two rates are about the
same on average. But the students of teachers who
ask less low-level and more high-level questions
ask more questions than the students of teachers
who ask more low-level and less high-level
questions. Susskind's study also includes a
category of competitive questions that have a
particularly negative impact on students' interest
in asking about anything. 

In addition, Susskind showed that a measurable
change in the classroom frequencies of
question-asking can occur as a result of the
following sequence of events (see Susskind's work
for important procedural details, such as
anonymity of records): 1) teachers make
predictions of the numbers of questions that will
be found in their classrooms, 2) researchers
measure what happens and demonstrate the results
to the teachers, 3) the results are discussed at a
few seminar meetings, at which the researchers and
teachers exchange ideas about the origin of
differences between expectations and facts. Such
discussions stimulate teachers to try different
approaches and improve the situation. After the
seminars, teachers are observed again. It is found
that on average the number and types of questions
teachers ask before and after the sequence of
events 1, 2, 3 do not differ much, although some
differences are discernible (for example, the
average number of questions asked by teachers
dropped by about 10\%, while the average
percentage of high-level questions increased from
about 50 by about 2 and the average percentage of
competitive questions decreased from about 8 by
about 2, with all these changes being comparable
with the magnitudes of corresponding standard
deviations but correlated in ways that
nevertheless allowed Susskind to draw his
conclusions). On the other hand, the observed
change in the behavior of students of the teachers
who went through the sequence 1, 2, 3, is
dramatic: the rate of their students'
question-asking doubles. Susskind's work warns
readers that there may be many reasons for why the
seminars could help teachers to change their
classroom behaviors and thus also change the
students' behavior (see also two paragraphs
below). The MOT suggests that when the human brain
sees a conflict between an assumption or
prediction it makes, especially about itself, and
actual results of observation and experiment, it
attempts to make changes in its space of knowledge
in order to become better prepared to deal with
new events (see Sections \ref{subsec:sto} and
\ref{subsec:hvtof}). It tries out new ideas. This
is precisely what Susskind says the teachers did
as a result of participation in the seminars.
Although no simple explanation is offered by
Susskind, the correlations he observed in the
study (see the original articles) still allow him
to make a few recommendations.

Susskind's recommendations for teachers are: to
reduce the number of questions teachers ask, to
ask questions of greater complexity, and to write
the best questions asked by students on the
blackboard. Furthermore, if teachers ask questions
about students' personal experiences, students
feel encouraged to ask questions concerning issues
that truly interest them.

Susskind also recommends videotaping of classrooms
as a means of creating a reliable source of
information for teachers about how they and their
students behave (the same recommendation is made
by many authors~\cite{CPEfilm}). Unfortunately,
studies of classroom regularities are rare (on the
order of 10 per 50 years before
Susskind's~\cite{Susskind1}, cf.~\cite{Lortie}).
Today's technology provides many means for
recording (including self-recording) and
discussing classroom regularities by the trainers
and trainees that are significantly better than
the ones that existed at the time of Susskind's
studies. The virtue of a film record is that a
trainee can see how his or her action appeared
from outside while the trainee also knows what
internal events in the trainee accompanied what
happened.

An example of observations that trainees need to
learn to make is that Susskind's study describes
only correlations among measured numbers of
various types of questions asked by teachers and
students (statistically analyzed assuming that
systematic errors cannot be large if various
observers agree in their counting and judgment of
the type of the questions). There is no claim of
observing or understanding causal relationships or
dynamical origins of the observed correlations,
even though researchers are tempted to think that
they see some causality~\cite{Susskind3}.

A description of what is observed must be
distinguished by trainees from an understanding of
the dynamical origin of what is observed. Using
the analogy of Bohr's model (see the beginning of
Section~\ref{sec:need}), one may say that the
description of the observed atomic spectra by
Bohr's model was an essential step on the way to
the discovery of the underlying dynamics of the
atom, but by itself did not provide an
understanding of this dynamics. Moreover, trainees
need to realize that understanding the dynamics of
behavior of teachers and students in interaction
is incomparably more difficult than just observing
and describing their behavior. This realization
can motivate the trainees to seek a better
understanding of the dynamics of their own
interactions with students.

Studies by Csikszentmihalyi et
al.~\cite{Teenagers} used modern remote
communication technology (such as beepers) to
monitor students' behavior during various
activities around the clock (the technique is
called experience sampling method). Results of
these studies agree with the model assumption
stated in Section~\ref{subsec:ed} and elaborated
on in Section~\ref{subsec:hvtof} that students are
strongly motivated by the visions they have of
their own future~\cite{future}. On the other hand,
the personal desire for and enjoyment of
participation in activities is associated with the
concept that Dewey described (in the context of
artistic expression) as follows~\cite{DeweyHenry}:
``Because of this wholeness of artistic activity,
because the entire personality comes into play,
artistic activity which is art itself is not an
indulgence but is refreshing and restorative, as
is always the wholeness that is health. There is
no inherent difference between fullness of
activity and artistic activity; the latter is one
with being alive.'' Csikszentmihalyi calls such
ultimate involvement in an activity
``flow''~\cite{Flow}, since one can say that a
person feels flowing fully immersed in the
activity and ceases to sense the flow of time. The
same phenomenon is known to exist in all
disciplines, including
science~\cite{HadamardFlow}, art, sport, and
teaching~\cite{teachingFlow}. The training of
teachers should include the experience of flow in
the process of teaching. Productive teaching is a
source of immense joy for a teacher who sees how a
student moves on in the process of learning, and
crosses otherwise insurmountable barriers. This
feature is heuristically included in the dynamical
part of MOT discussed in
Section~\ref{subsec:hvtof}.

The training of teachers must include basic issues
of communication and cooperation with parents.
Numerous examples illustrate that teachers and
parents must communicate~\cite{parents}. Parents
are major players in the student's
life~\cite{teenagersFamily}. They can help a
teacher understand where the student is coming
from and how to help a student in getting on the
path of learning, which are indispensable elements
in productive teaching according to the MOT. On
the other hand, the model suggests that
communication between teachers and parents as
observers of the same reality from different
frames of reference encounters the same type of
difficulties as communication between teachers and
students. Of course, the case of teachers and
parents involves significant internal events (most
parents deeply care about their children) and
needs an extended meta-dictionary (for parents,
the subject matter is of secondary importance to
how their children fare). The MOT predicts and
explains why a clear communication in these
circumstances must be difficult and requires
training, see Section~\ref{subsec:cie}.

The degree of success in training of a pool of
teachers in productive teaching will depend on the
selection of candidates~\cite{selection}. Ideally,
a selection process should be based on strengths
that the candidates exhibit in their records and
during ``auditions'' in the role of a teacher. As
in the case of good schools of music, where it is
not enough to just know how to hold an instrument
and play a sequence of notes, so in the case of
teaching it is not enough to know something about
the subject matter and spell it out in front of
the class (e.g., see the MOT predictions
concerning difficulties in communication about
internal events described in
Section~\ref{subsec:cie}). The analogy with music
makes it clear that people who start learning how
to teach early, for example, in scouts, can
develop their skills far beyond average before
they come to the ``audition'' appropriate for
their admission to the school of productive
teaching as defined by the MOT.

If someday training in productive teaching becomes
a regularity of the educational system, one can
imagine that promising students will be immersed
in the contexts of productive learning by giving
them responsible teaching and administrative roles
in their schools and they will be helped in
practicing their skills as part of their work. The
help they may provide in return in teaching
younger students and running the school could in
principle reduce the burdens on teachers and
provide the teachers with more time than they have
now for their own professional
development~\cite{studentteachers}. The MOT
provides physical arguments for the necessity of
increase in the amount of time made available to
teachers for training.

It is clear at this point that the heuristic
concept of productive teaching defined by the MOT
implies a long list of suggestions for training of
teachers and a corresponding list of questions
that require answers. With current capabilities
for research on teaching and learning it should be
possible to begin systematic studies on the
applicability of the concepts of productive
teaching and training in practice.

It is expected that a trainee teacher who learns
the concept of productive teaching will
voluntarily express gratitude to a trainer for
providing the lesson. The reason is that if the
trainer works on the lesson and the trainee
understands what the result is supposed to be, it
will be clear to the trainee familiar with the MOT
that confirmation of the result is what the
trainer needs to obtain in order to know the
degree to which the task is accomplished.
Observing how the trainee behaves in contact with
students under supervision (or aware of being
observed or filmed) is not sufficient because a
trainee may behave like an actor, who knows how to
follow instructions given by a director and
perfectly fakes a character while not actually
becoming one. Since the concept of productive
teaching is learned inside the trainee, only the
trainee can provide the information that the
lesson is learned. There is no point in asking the
trainee if he or she has understood the concept
before it is observed as emitted by the trainee.
When it occurs, the information is provided by the
trainee in a voluntary expression of appreciation
of the concept (the concept includes the
appreciation of the work of the trainer and
understanding by the trainee that the voluntary
expression of this appreciation is the way by
which the trainer knows the result). Honest,
self-motivated discussion of the concept cannot be
faked since it can only be based on concrete
events in which the trainee behaves in a sincere
way and thus learns from them. If that is not the
case, the trainee only pretends comprehension of
the concept. Attempts to hide confusion and
maintain superficial claims to understanding are
not difficult to identify. When the trainer points
out a problem related to the trainee's conception
of productive teaching, the trainee's reaction is
itself a measure of the depth of understanding of
the concept. So, it is relatively easy to know
whether a trainee is still in the woods. But it is
not possible to tell that the training is
completed until the trainee voluntarily provides
the evidence. 

Everyone who works in the capacity of a teacher
does so in large part as a result of prior
training. If that training was not based on the
principles of communication between brains that
are identified in the model, the principles may
remain unknown to the person. It is then not the
person's fault that she or he is not able to
create a productive teaching environment for
learning by others. Such cases can be gradually
eliminated. According to the MOT, the elimination
requires an extension of the concept of teaching
from effective to productive. Those who are to
train teachers would have to demonstrate
understanding of the difference before becoming
trainers.

\section{ Conclusion }
\label{sec:c}

The MOT proposed here provides a description of
the concepts of effective and productive teaching
that are not at all new in their meanings. Authors
of great influence on our thinking have discussed
these meanings extensively. The original aspect of
the model is its simplicity. Namely, it sketches a
picture of these concepts in simple physical
terms. 

This is useful for seeking systemic solutions to
problems in Education since one has to define the
goals of the system before one attempts to figure
out how to create systemic conditions in which
these goals can be achieved. It is clear that the
system must be designed in such a way that the
forces pushing toward the goal are not
counterbalanced by other forces that might also
act in a system and win if the system is designed
without the necessary understanding. If the goal
is productive teaching defined in the MOT, the
system forces must be arranged in such a
configuration that the basic forces that drive
productive teaching can function and perform their
work. 

However, the major problem of figuring out a
suitable arrangement is not solved
here~\cite{laststep}. For example, the question of
how to train people who are to legislate,
administer, and judge the work of teachers is not
resolved. This problem can be called the system
government problem~\cite{systembrain}. It is clear
that it depends on the competence of the system
government if teachers (appropriately trained) can
perform in agreement with their training in the
conditions created for their work (see
Section~\ref{subsec:tet}). The system government
problem also means that the search for systemic
solutions cannot be limited to the educational
system alone. The bedrock assumption is that in
all circumstances where teaching occurs in a
system, both teacher and student are learning and
they learn from each other in identical ways as
observers who exchange information about the same
reality.

The insights the model offers regarding effective
teaching (Section~\ref{sec:kot}) and productive
teaching (Section~\ref{sec:dot}) help in
comprehending Sarason's early prediction of the
failure of educational
reforms~\cite{SarasonMemoir}. Sarason's prediction
is based on his concept of a context of productive
learning. Until now, there existed no simple
description of what this concept means. According
to the model, a context of productive learning
exists for a student in the process of productive
teaching (see Section~\ref{subsec:pt}). A reform
effort cannot be successful in creating contexts
of productive learning for students unless the
processes of productive teaching are center stage
in the system. Thus, if an expert looks at a
reform plan, including plans to monitor and
document the performance of the system in the long
run, and sees that nowhere in it the key processes
are paid due attention, the prediction of failure
may be made long before the reform is implemented.
According to Sarason, Bensman~\cite{Bensman} and
Levine~\cite{Levine} provided credible
descriptions of reform efforts that come the
closest to incorporating contexts of productive
learning in school practice.

The model picture also sheds some light on the
belief that a rigorous scheme for training
teachers in science (and through them their
students) can by itself stimulate interest in
learning. The model suggests instead that the
kinematical image of the process of science must
result from its proper dynamics (whatever it may
be). When the image is artificially created
through special constraints and incentives,
neither teachers nor students can learn what the
relevant forces are (whatever they may be). A less
arbitrary alternative for the design of reforms is
to secure facilitation of action of the natural
forces of productive teaching: strengths of
teachers and students. For this purpose, the
knowledge of teachers needs to include the
meta-dictionary of teaching. The model provides a
preliminary definition of the meta-dictionary,
showing that a dictionary that is limited to
subject matter alone is insufficient to discuss
educational reform.

When teachers and students are seen as observers
of the world from different frames of reference
and the kinematics and dynamics of their
communication about what they observe are analyzed
in simple terms of a physical model, it becomes
clear that the interpersonal relationship between
a teacher and a student is the key to productive
teaching defined in the MOT. According to the
model, productive teaching is not only effective
in terms of mastery of the subject of study. It
also helps a student to understand the role the
teacher plays in the process, a facilitator of
learning rooted in the individual's strengths.
Through appreciation of this role, a student
becomes a person who understands the principles
and virtues of communication between people. This
is why the model concept of productive teaching
appears to provide a picture of what systemic
solutions need to support in order to make a
transition to a society in which words and ideas
can help people in carrying out the work that is
necessary for their well-being~\cite{Drucker}. 

However, the model also says that the number of
dimensions in the space of events registered in a
brain is large (see
Appendix~\ref{app:dimensions}). Therefore, the
corresponding relationships between frames of
reference of different brains are complex,
nonlinear, and topologically nontrivial. MOT
predicts through these circumstances that an
exchange of sequences of coordinates without
paying much attention to their proper
interpretation in different frames of reference
may be easily mistaken for the concept of
communication between brains. But if teaching is
equated to handing down to a student a sequence of
coordinates $t$ formed using a language known in
fact only to the teacher, and then checking if the
sequence $t$ is correctly handed back, not only is
the teaching ineffective and students unfairly
judged, but also a new phenomenon is created.
Every sequence $t$ can be handed around without
clarity or need for clarity of its meaning in
terms of real events. In such case, everybody can
``teach'' because there is no need to know either
the content or how to transfer the content. The
model predicts in simple terms that teaching
cannot be improved by streamlining such a mindless
process of handing down statements without
content. The prediction holds no matter how much
money may be spent on such a process because the
process has nothing to do with productive teaching
defined in the model. Thus, the model supports a
claim that this process will not lead to
transition toward new forms of society such as the
one envisioned by Drucker~\cite{Drucker}.

According to the MOT, productive teaching demands
preparation of teachers who can communicate with
students in many more dimensions than only those
of the subject matter (Section~\ref{subsec:tipt}).
Such advanced preparation requires an
apprenticeship system in which novices are taken
care of by mentors over extended periods of time
and may stay in touch with the mentors during
their subsequent professional activities.
Certainly, such preparation of teachers resembles
processes of induction or mediated entry in
professional disciplines and as such implies an
open-ended self-correcting evolutionary process of
change in the discipline of teaching based on
ongoing research and development of the highest
standards. By comparison with contemporary systems
of education, such an evolutionary process of
change demands new organization because the
existing organization is focused too much on
handing down sequences of words without paying due
attention to their meanings. This illustrates how
potentially far-reaching predictive power physical
models of teaching may have. 

Understood as a process based on interaction
between different brains, teaching cannot be
described by models that are limited to learning
by only one brain and do not involve the transfer
and processing of information in interaction
between different brains. However, models of how a
single brain learns are very important for
teaching despite their
limitations~\cite{Hebb,Marr,Hopfield1,
SejnowskiPaulsen}. For example, consider that the
process of gene expression in the formation of an
individual's memory~\cite{Bailey,Udo} (and perhaps
also in other Hebbian processes) depends on
factors such as attention~\cite{KentrosAttention}.
When combined with the MOT, this consideration
suggests that an educational system harms the
development of students' brains if productive
teaching is rare and students are regularly
prevented from focusing their attention on
productive learning. Given the model of teaching,
such an educational system can no longer be
considered legitimate. In order to become
legitimate, it has to engage in a long-term
research and development process of the highest
quality on an appropriate scale, concerning
productive teaching and its implementation on a
regular basis.

The model described here may be seen as an attempt
to incorporate a number of ideas that originate in
various arts and sciences into a
physically-motivated structure. Such attempts are
in need of building bridges between different
disciplines; for example, between psychology and
physics in terms of neural science. It is
encouraging that physicists and neural scientists
are already involved in
discussions~\cite{BorgGraham,Hopfield2,
PhysicsNeuroscience}, and that neural scientists
appear to be in communication with
psychologists~\cite{NeuralPrinciples}. Even though
a physical MOT may in principle offer a framework
in which all required elements have a chance to
eventually find their place, it is predicted that
much more precise physical models of teaching will
be needed than the very preliminary one described
here. Nevertheless, because this preliminary MOT
has a simple physical structure with many concrete
implications regarding otherwise complex issues of
teaching, it can be predicted on general
grounds~\cite{Rogers} that physical MOTs will be
useful in guiding reform of educational systems.

\vskip.05in
{\bf Acknowledgments} 
The author would like to thank Seymour B. Sarason 
and Kenneth G. Wilson for many discussions.

\begin{appendix}

\section{ Space of events registered in a brain }
\label{app:dimensions}

The word ``event'' is associated in physics with
what happens in a set of points in space-time. But
a learning human brain obtains information about
events only in the form of input it can receive. A
heuristic idea of parameterization of the
information about events registered in a brain is
used below to define and visualize the concept of
space of events that is used in
Sections~\ref{sec:kot} and~\ref{sec:dot}.

Consider how a snapshot of a learning individual's
environment forms a record of an event and how
such a record is received as input in the
individual's brain. The input comes from the
visual system and causes changes in the brain.
Complete description of the changes undoubtedly
requires a large set of parameters. But even
without knowing the appropriate set of parameters
and their values one may postulate that the
changes ultimately result from the interplay of
three elements: 1) The light actually received in
the eyes, 2) performance of the system of vision
that provides input to the brain, and 3) the
activity in the brain at the moment of reception
of the input from the visual system.

The first two elements contribute to the process
of registration of events by the individual's
brain in familiar ways. For example, a student may
be observing what happens on a laboratory table
and, perhaps, must be wearing corrective lenses in
order to see with sufficient precision. The third
element, i.e., the ongoing brain activity whose
change constitutes registration of an event, is
least well-understood. Obviously, such brain
activity is not based solely on the momentary
input from the visual system. It depends also on
the input from all other senses and from a large
number of body organs. Ultimately, the processes
that occur in the brain itself determine how
events are registered. Factors such as attention
and interest (or lack thereof for various, complex
reasons) play significant roles. Also, since the
brain develops over time, its activity at every
moment depends on its biophysical history and the
current structure this history has produced,
including memory. It is clear that the brain's
reaction to every momentary input it receives is
not a simple function of this input. Thus, the
concept of space of events registered in a brain
requires a definition.

It is postulated that at every moment in time a
brain located at any place in space consists of
particles and fields in some state that in
principle may be described using a very large
number of physical parameters. A considerably
smaller number of parameters describes the
biophysical state of the brain, which corresponds
to the underlying state of particles and fields.
These parameters change over time. A change in
this set of parameters over a short period will be
called a raw representation of an event in the
brain. The length of the suitable period is of
secondary importance. For the MOT,
periods on the order of one hundredth of a second
are perhaps appropriate since a human eye cannot
discern images that appear with frequency larger
than on the order of one hundred per second, ears
cannot discern words spoken faster than about ten
per second, typical muscular reaction time is on
the order of one fifth of a second, etc. 

The raw representations of events are somehow
proc-essed in a brain so that the results can be
stored in memory, recalled, and recognized. These
results are called events registered in a brain
and, by definition, they are elements of the space
of events registered in the brain. 

The space of events registered in a brain is
certainly not isomorphic with the space of
physical events outside the volume swept in
space-time by the brain because the events
registered in the brain depend on processes that
involve physical events inside this volume. The
latter are specific to the brain in which they
happen. It is clear that events registered in
different brains cannot be compared in any simple
way.

Studies of the nervous systems of simple
organisms~\cite{Kandel,KentrosAttention,Udo,
Bailey} indicate how the raw representations of
events may be processed and how the results of
this processing may be registered in an organism's
brain. The human brain is so complex that attempts
to fully explain how it works are unlikely to
succeed in the near future~\cite{Marr}.
Nevertheless, one may imagine the space of events
registered in a human brain using the concept of
coordinates. This coordinate picture is referred
to in Sections~\ref{sec:kot} and~\ref{sec:dot} and
helps in appreciation of the magnitude of
dimensionality of the events that underlie the
process of productive teaching discussed in
Section~\ref{sec:dot}. 

In order to introduce the relevant coordinates,
one may start from the postulate that the human
brain is built from units that are connected by
links. The smallest unit to think about would be a
neuron and the smallest link would be a synapse. A
more suitable level of analysis is that neurons
work in groups and these groups may be considered
units that are connected by complex links. A
candidate to consider for a unit would be Hebb's
assembly~\cite{Hebb}. Fortunately for the purposes
of the model, one may postulate the existence of
units and links without specifying precisely their
nature, except that there are many units and many
more links among them. The physical arrangement of
units and links in the brain, such as their
apparent spatial extent or overlap, is irrelevant.

In order to quickly imagine the coordinates of
events, it is best to consider first only two
units connected by just one link. It is postulated
that the magnitude of activity in the link at any
moment and any position of the brain can be
described by one number. This number is treated as
a coordinate of content of the event registered in
the brain. Note that the coordinate most likely
describes a compact dimension since the link
ceases to function properly and is turned off when
the activity increases above a certain threshold.
Thus, a coordinate above threshold does not exist
and the threshold value is considered equivalent
to zero~\cite{compact}.

Since the brain is considered to be built from
many units and the number of just pair-wise links
between $m$ units, denoted by $n$, is much larger
than $m$, $n$ = $m(m-1)/2$, the content of an
event registered in the brain can be identified
with an element in an $n$-dimensional space with a
large $n$. It is postulated that $n$ is fixed and
the same for all brains. More complex alternatives
do not need to be discussed here. Including the
time and place of registration, the number of
dimensions in the space of events registered in a
brain is $N = 4 + n$, a much larger number than 4
that applies only to space-time. The magnitude of
$N$ suggests greater difficulties with precise
communication between observers about events their
brains register than the observers typically
encounter already in communication about
point-like events in four-dimensional spaces.

\end{appendix}


\begin{thebibliography}{99}
\bibitem{wordsandconcepts} 
Words are used to describe and to name concepts.
\bibitem{externaldictionary}
Building one's own ways of thinking is a different
process from the process of learning an external
dictionary of others, but one cannot undertake 
studies that lead to creation of genuinely new 
ways of thinking without sufficient information 
obtained from others, see Section~\ref{subsec:hvtof}.
\bibitem{classical} 
The procedure is classical, not quantum.
\bibitem{learning} 
See Ref.~\cite{Sarason} for examples of meanings
people associate with the word {\it learning} that
has the highest count in the literature on
education.
\bibitem{inadequate} 
The adequacy of rules that are used in setting
proficiency levels (for example, how they are
adjusted to the development of students' brains in
time) is not publicly discussed.
\bibitem{testing}
This problem is not solved by testing.
\bibitem{internal}
Whether only $T$ or both $T$ and $S$ know that it
happens, attempts at communication about internal
events create for them a common meta-dictionary of
teaching~\cite{GlazekSarason}.
\bibitem{SarasonCPL2} 
See Ref.~\cite{SarasonCulture}, Ch. 21, and 
trace back to Ref.~\cite{SarasonCPL1}, where 
the school culture is identified as one of the 
obstacles to change in the classroom; then see 
Refs.~\cite{Sarason} and \cite{GlazekSarason}.
\bibitem{Tsystem} 
The process of teaching cannot be understood 
here as limited to what $T$ does because $T$
works in a system and the system may dictate
norms that generate anxiety. E.g., see Ref.~\cite{Cutts}.
\bibitem{quantum}
Physical laws lead to the formation of the 
genetic code and its expression in ways that
may be not understandable using classical 
concepts, because these concepts do not 
apply in description of quantum phenomena.
\bibitem{withoutT} 
In rare cases, individuals are able to learn using
information from others almost entirely without
direct personal contact (e.g., only reading books), 
and the teaching they receive this way may be
effective in the sense of learning a specialty.
Productive teaching includes personal contacts in
order to avoid unproductive consequences for the
learner of learning almost entirely without
personal contacts.
\bibitem{Wieman1} 
Wieman's observation that teaching may be not 
as productive for students as a teacher
thinks~\cite{Wieman}, appears related to a
possibility that it is difficult for a person
to pay attention to issues of teaching when the
person needs to focus on the subject matter.
\bibitem{selfcorrectioninmath}
The same need for self-correction based on
comprehension is evident in students of physics
who mechanically transform equations without
realizing what they are doing and cannot see
and correct errors they make.
\bibitem{teachingFlow}
Ref.~\cite{Teenagers}, Chap. 9. The chapter is
focused on teachers providing conditions for flow
to students and suggests that teachers can do it
on the basis of their own experience of flow.
\bibitem{systembrain}
Perhaps a better name is ``the system brain
problem,'' since it reflects the need for 
learning by a complex system that is to 
evolve consciously.
\bibitem{compact}
On a circle, as example of a compact dimension, an
angle can be chosen as a co-ordinate: it ranges
from 0 to $2\pi$, and the point with co-ordinate
$2\pi$ is the same as the one with co-ordinate 0.
\end{thebibliography}
\end{document}